\documentclass[sigconf]{acmart}

\usepackage{booktabs}
\usepackage{algorithmic}
\usepackage{graphicx}
\usepackage{textcomp}
\usepackage{xcolor}
\usepackage{tikz}
\usetikzlibrary{trees}
\usepackage{adjustbox}
\usepackage{xspace}
\usepackage{soul}
\usepackage{pgf-pie}
\usepackage{enumitem}
\usepackage{tcolorbox}
\usepackage{subcaption}
\usepackage{wrapfig}
\usepackage{multirow} 
\usepackage{pgfplots}
\usetikzlibrary{decorations.pathreplacing}

\newcommand{\example}[2]{
\vspace{-2mm}
\begin{tcolorbox}[colback=gray!10,colframe=white,width=0.96\columnwidth,arc=0.3mm, auto outer arc,boxrule=0.1pt,boxsep=1pt,top=0pt,bottom=0pt,left=2pt,right=2pt]
\textbf{\textit{E.g.,}} \textit{``#1''}

{\small\hfill --- \textit{by #2}}
\end{tcolorbox}
\vspace{-2mm}
}

\newcommand{\inlinecode}[1]{\texttt{\small #1}\xspace}

\usetikzlibrary{trees, shapes, arrows.meta, positioning}
\usetikzlibrary{matrix, arrows.meta, positioning}
\usetikzlibrary{positioning,calc,decorations.pathreplacing}

\definecolor{lightred}{RGB}{255,182,193}    
\definecolor{lightgreen}{RGB}{144,238,144}
\definecolor{lightyellow}{RGB}{255,255,153}

\newcommand{\hlred}[1]{\sethlcolor{lightred}\hl{#1}}
\newcommand{\hlgreen}[1]{\sethlcolor{lightgreen}\hl{#1}}
\newcommand{\hlyellow}[1]{\sethlcolor{orange!30}\hl{#1}}

\definecolor{color1}{RGB}{220, 38, 38}   
\definecolor{color2}{RGB}{249, 115, 22}  
\definecolor{color3}{RGB}{234, 179, 8}   
\definecolor{color4}{RGB}{34, 197, 94}   
\definecolor{color5}{RGB}{59, 130, 246}  
\definecolor{color6}{RGB}{168, 85, 247}  %

\newcounter{finding}
\newcommand{\finding}[1]{\refstepcounter{finding}
	\begin{center}
		\begin{tcolorbox}[colback=gray!10,colframe=black!50,width=1\columnwidth,arc=1mm, auto outer arc,boxrule=0.5pt,boxsep=5pt,left=3pt,right=3pt,top=0pt,bottom=0pt]
		\textbf{Finding \arabic{finding}:} #1
		\end{tcolorbox}
	\end{center}
}

\newcommand{\TaskIdentification}{TSK\xspace}
\newcommand{\ContextIdentification}{CTX\xspace}
\newcommand{\ConstraintIdentification}{CST\xspace}
\newcommand{\KnowledgeRecall}{KRL\xspace}
\newcommand{\ControlFlowConstruction}{CFL\xspace}
\newcommand{\ApproachComparison}{CMP\xspace}
\newcommand{\AmbiguityRecognition}{AMB\xspace}
\newcommand{\ScaffoldCodeGeneration}{SCG\xspace}
\newcommand{\CompleteCodeGeneration}{CCG\xspace}
\newcommand{\UnitTestCreation}{UTC\xspace}
\newcommand{\PostHocAlternativeExploration}{ALT\xspace}
\newcommand{\EdgeCaseIdentification}{EGC\xspace}
\newcommand{\FlawIdentification}{FLW\xspace}
\newcommand{\StyleCheck}{STY\xspace}
\newcommand{\SelfAssertion}{SFA\xspace}
\newcommand{\Placenameone}{CONT\xspace}
\newcommand{\Placenametwo}{GUIDE}

\begin{document}

\title{A Study on Thinking Patterns of Large Reasoning Models in Code Generation}

\author{Kevin Halim}
\email{kevincon001@e.ntu.edu.sg}
\affiliation{%
  \institution{Nanyang Technological University}
  \country{Singapore}
}

\author{Sin G. Teo}
\email{teosg@i2r.a-star.edu.sg}
\affiliation{%
  \institution{Institute for Infocomm Research (I2R), A*Star}
  \country{Singapore}
}

\author{Ruitao Feng}
\email{ruitao.feng@scu.edu.au}
\affiliation{%
  \institution{Southern Cross University}
  \country{Australia}
}

\author{Zhenpeng Chen}
\email{zhenpeng.chen@ntu.edu.sg}
\affiliation{%
  \institution{Nanyang Technological University}
  \country{Singapore}
}

\author{Yang Gu}
\email{yang.gu24@u.nus.edu}
\affiliation{%
  \institution{National University of Singapore}
  \country{Singapore}
}

\author{Chong Wang}
\email{chong.wang@ntu.edu.sg}
\affiliation{%
  \institution{Nanyang Technological University}
  \country{Singapore}
}

\author{Yang Liu}
\email{yangliu@ntu.edu.sg}
\affiliation{%
  \institution{Nanyang Technological University}
  \country{Singapore}
}

\begin{abstract}


Currently, many large language models (LLMs) are utilized for software engineering tasks such as code generation. The emergence of more advanced models known as large reasoning models (LRMs), such as OpenAI's o3, DeepSeek R1, and Qwen3. They have demonstrated the capability of performing multi-step reasoning. Despite the advancement in LRMs, little attention has been paid to systematically analyzing the reasoning patterns these models exhibit and how such patterns influence the generated code.
This paper presents a comprehensive study aimed at investigating and uncovering the reasoning behavior of LRMs during code generation. We prompted several state-of-the-art LRMs of varying sizes with code generation tasks and applied open coding to manually annotate the reasoning traces. From this analysis, we derive a taxonomy of LRM reasoning behaviors, encompassing 15 reasoning actions across four phases.

Our empirical study based on the taxonomy reveals a series of findings. First, we identify common reasoning patterns, showing that LRMs generally follow a human-like coding workflow, with more complex tasks eliciting additional actions such as scaffolding, flaw detection, and style checks. Second, we compare reasoning across models, finding that Qwen3 exhibits iterative reasoning while DeepSeek-R1-7B follows a more linear, waterfall-like approach. Third, we analyze the relationship between reasoning and code correctness, showing that actions such as unit test creation and scaffold generation strongly support functional outcomes, with LRMs adapting strategies based on task context. Finally, we evaluate lightweight prompting strategies informed by these findings, demonstrating the potential of context- and reasoning-oriented prompts to improve LRM-generated code. Our results offer insights and practical implications for advancing automatic code generation.
\end{abstract}

\ccsdesc[500]{Software and its engineering}

\keywords{Large Reasoning Models, Code Generation, Taxonomy}

\received{20 February 2007}
\received[revised]{12 March 2009}
\received[accepted]{5 June 2009}

\maketitle

\section{Introduction}

Large language models (LLMs) have become increasingly prominent in software engineering, particularly in the domain of automated code generation. Tools like GitHub Copilot \cite{Friedman_2022} demonstrate the utility of LLMs by producing code that is often syntactically correct and contextually plausible \cite{austin2021program}. These models assist with tasks such as code completion, code summarization, and bug fixing, offering substantial productivity gains. However, LLM-generated code often lacks deeper semantic understanding, leading to issues such as functional requirement violations \cite{zhang2025llm}, hallucinations \cite{ji2023survey}, and outputs that do not align with user intent. Additionally, their ``black-box'' nature \cite{zhao2024explainability} and limited capacity for robust reasoning pose serious challenges to their reliability in practical software development.

\begin{figure}
    \includegraphics[width=1\linewidth]{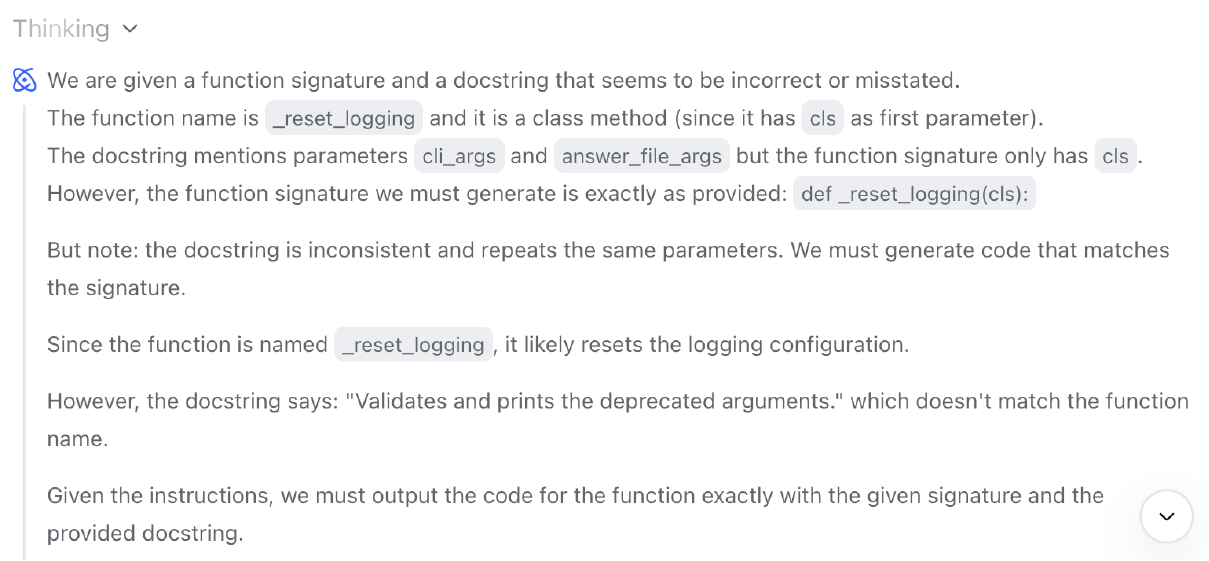}
    \caption{An motivating example where DeepSeek-R1 is thinking before generating code.}
    \label{fig:trace}
\end{figure}

The emergence of Large Reasoning Models (LRMs) represents a significant advancement in model reasoning capabilities. Notable examples include OpenAI's \texttt{o}3 \cite{openai}, Anthropic Claude 4 \cite{claude}, Qwen3 \cite{qwen3}, and QwQ \cite{qwq32b}. These models generate explicit intermediate reasoning steps, commonly called \textit{reasoning traces}
, which offer valuable insight into how outputs are produced \cite{marjanovic2025deepseek}. In contrast to traditional LLMs that primarily rely on pattern matching, LRMs are designed for multi-step deliberation and better generalization to complex problems \cite{bestaReasoningLanguageModels2025}. This enhanced reasoning improves effectiveness and also plays a crucial role in transparency and interpretability, fostering understanding, building user trust, and ensuring alignment in application tasks. For instance, as illustrated in Figure~\ref{fig:trace}, DeepSeek-R1 identifies a given signature as a class method based on the \inlinecode{cls} parameter and detects ambiguities in the signature and docstring during its internal reasoning before generating code.

On the other hand, to better understand the reasoning capabilities of language models, recent research has proposed various taxonomies describing how models reason across different domains. Several studies \cite{marjanovic2025deepseek,ming2025helpful,plaat2024reasoning,bandyopadhyay2025thinking} focus on general problem-solving, critique, mathematical reasoning, and structured reasoning phases. While these taxonomies offer valuable insights, they are not tailored to code-related tasks, limiting their relevance to code generation. In contrast, other works \cite{wei2025evaluating,liu2024codemind} examine reasoning in code generation, highlighting issues in logical consistency and control flow. However, these studies evaluate traditional LLMs and do not investigate the reasoning traces (as shown Figure~\ref{fig:trace}) of large reasoning models (LRMs). As a result, existing literature either overlooks code generation or focuses on models not designed for explicit reasoning traces.  This raises an important question: \textit{How do LRMs perform reasoning in code generation, and are there identifiable patterns in their reasoning behaviors?}


\textbf{Taxonomy.} To bridge this gap, we analyze reasoning traces of LRMs in code generation, aiming to uncover and model common reasoning behaviors in how models interpret prompts and generate code, thereby offering insights to improve code generation techniques. We focus on five open-source Qwen-series LRMs, including DeepSeek-R1-7B, Qwen3-(1.7B, 8B, 14B), and QwQ-32B, evaluated on Python tasks from the CoderEval benchmark~\cite{yu2024codereval} using prompts designed to elicit explicit reasoning. To identify reasoning behaviors, we apply an open coding method to manually annotate 1,150 traces across different task complexities from the five models, noting recurring reasoning actions. We organize these behaviors into four phases, merging and refining similar actions to construct a taxonomy of 15 reasoning actions, spanning from requirements gathering to reflection.

\textbf{Study.} Building on this taxonomy, we conduct an empirical study to address a series of research questions.
\textbf{RQ1:} We identify common reasoning patterns in LRM-based code generation by analyzing the sequence of reasoning actions within traces. The results show that LRMs generally follow a human-like coding workflow—analyzing requirements, clarifying ambiguities, comparing solutions, implementing code, and reviewing for defects—while simpler tasks involve lighter reasoning, and more complex tasks trigger additional actions such as scaffolding, flaw detection, or style checks.
\textbf{RQ2:} We compare reasoning behaviors across different LRMs. The results indicate that Qwen3 models exhibit highly similar, iterative reasoning patterns, whereas DeepSeek-R1-7B follows a more linear, waterfall-like approach, likely reflecting differences in the reasoning traces used during training.
\textbf{RQ3:} We analyze how reasoning behaviors impact functional correctness by correlating Pass@1 with reasoning actions and patterns. Unit test creation and complete code generation strongly support correctness, and LRMs adjust their reasoning strategies such as knowledge recall, alternative exploration, scaffold code generation, and edge case generation based on task dependency and context.
\textbf{RQ4:} We evaluate the feasibility of two lightweight prompting strategies as potential improvements over the initial prompt, motivated by the key findings. The results highlight the potential of incorporating context or reasoning-oriented guidelines into prompts to enhance LRM-generated code.
Beyond the empirical results, we also derive practical implications for researchers and developers to facilitate the advancement and application of LRM code generation.

The contributions of this paper are as follows:
\begin{itemize}[leftmargin=20pt]
    \item We develop a comprehensive taxonomy of 15 reasoning actions across 4 phases in LRM-based code generation.
    \item We conduct an empirical study that uncovers common reasoning patterns, highlights differences in reasoning behaviors across models, analyzes the impact of reasoning behaviors on functional correctness, and evaluates the feasibility of lightweight prompting improvements.
    \item We release a dataset of 1,150 annotated reasoning traces to support future research on the reasoning capabilities of LRMs.
\end{itemize}

\section{Study Setup}


\begin{figure*}
    \centering
    \includegraphics[width=0.8\textwidth]{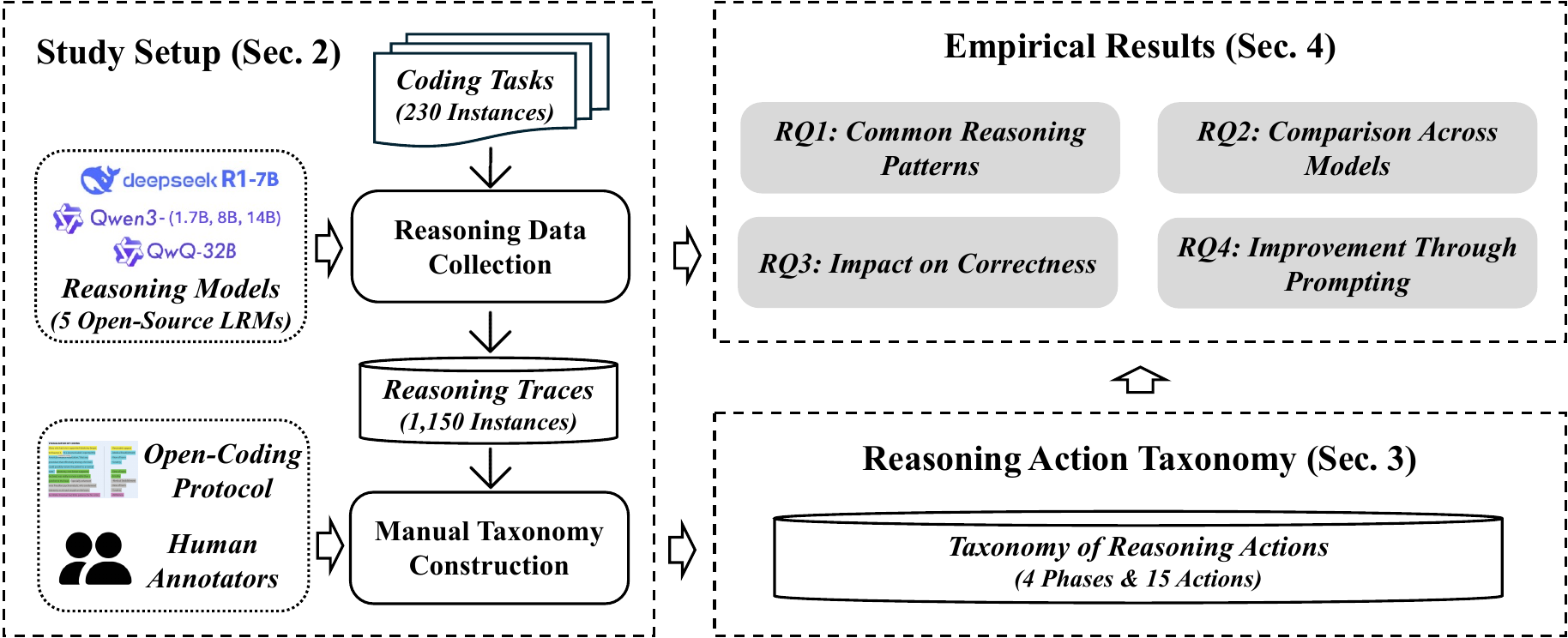}
    \caption{Methodology overview of our study.}
    \label{fig:methodology}
\end{figure*}

Figure \ref{fig:methodology} presents an overview of the methodology employed in this study. We begin by prompting five selected large reasoning models (LRMs) to generate reasoning traces on 230 code generation tasks. Two human annotators then apply open coding methodology on the collected $230 \times 5 = 1,150$ reasoning traces to manually construct a taxonomy of reasoning behaviors exhibited by LRMs, resulting in 15 distinct reasoning actions organized across 4 phases.

Building on this taxonomy, we conduct an empirical study to address a series of research questions. First, we examine the common patterns of action combinations and analyze the rationales underlying these reasoning patterns (\textbf{RQ1}). Next, we compare different LRMs in terms of the reasoning actions and patterns they employ, highlighting similarities and discrepancies (\textbf{RQ2}). We then investigate how reasoning behaviors influence the functional correctness of generated code and distill key insights (\textbf{RQ3}). Finally, we explore the feasibility of lightweight prompting-based strategies for improving LRMs and share the lessons learned (\textbf{RQ4}). The research questions are listed as follows:
\begin{itemize}[leftmargin=20pt]
    \item \textbf{RQ1 (Common Reasoning Patterns)}: How do LRMs combine and perform individual reasoning actions during code generation?
    \item \textbf{RQ2 (Comparison Across Models)}: To what extent do different LRMs exhibit similar or divergent reasoning behaviors?
    \item \textbf{RQ3 (Impact on Correctness)}: How do the identified reasoning behaviors affect the functional correctness of the generated code, and what key insights can be drawn from the analysis?
    \item \textbf{RQ4 (Improvement Through Prompting)}: How can prompting strategies informed by these behaviors enhance the effectiveness of LRMs?
\end{itemize}

\vspace{-3mm}
\subsection{LRM Selection}
We select target reasoning models according to the following criteria: First, their internal reasoning traces must be accessible. This excludes certain commercial models, such as OpenAI’s \texttt{o}-series, which do not provide public API access to such traces. Second, the selected models should span a range of different sizes.
Based on these criteria, we choose the following open-source models.

\begin{itemize}[leftmargin=20pt]
    \item \textbf{DeepSeek-R1-7B}~\cite{deepseekai2025deepseekr1incentivizingreasoningcapability}: A large reasoning model developed by DeepSeek AI based on the Qwen2.5-Math-7B model. Despite its relatively small size, it demonstrates competitive performance comparable to that of larger models.
    
    \item \textbf{Qwen3-(1.7B, 8B, 14B)}~\cite{qwen3}: A family of reasoning-focused models developed by Alibaba's Qwen team, offering significant improvements over prior versions such as Qwen2.5 and QwQ. This study employs the 1.7B, 8B, and 14B variants to represent lightweight, small, and medium-large sizes, respectively.
    
    \item \textbf{QwQ-32B}~\cite{qwq32b}: A 32-billion-parameter reasoning model developed by Alibaba's Qwen team, trained using supervised fine-tuning followed by reinforcement learning. QwQ demonstrates strong reasoning capabilities, achieving performance on par with state-of-the-art models.
\end{itemize}

To better reflect real-world usage and explore the model's natural reasoning patterns, we adopt the default configuration (temperature is 0.6) as recommended by model vendors~\cite{deepseek_hf} and typically used by end users. 

\subsection{Reasoning Data Collection}
To collect reasoning data, we need to determine the code generation dataset on which to run the LRMs. We apply the following criteria for dataset selection: First, it should be constructed from realistic code repositories and widely adopted in prior studies; Second, it should cover code generation tasks that involve both standalone and non-standalone settings; Third, it should categorize tasks into different levels of difficulty. Based on these criteria, we selected CoderEval~\cite{yu2024codereval} from a pool of widely used benchmarks such as HumanEval~\cite{chen2021evaluating}, MBPP~\cite{austin2021program}, ClassEval~\cite{du2023classeval}, and DevEval~\cite{li2024deveval}. 

CoderEval is specifically designed to emulate real-world code generation challenges and is a comprehensive benchmark that simultaneously incorporates dependency-aware tasks and multi-level difficulty settings. It comprises 230 tasks, each for Python and Java, ranging from \textit{self-contained} snippets to \textit{project-runnable} tasks requiring cross-file dependencies:
\begin{itemize}[leftmargin=20pt]
\item \textbf{\textit{self-contained}}: Fully isolated tasks that can be executed without any dependencies.
\item \textbf{\textit{slib-runnable}}: Tasks requiring only standard libraries, with no need for additional packages.
\item \textbf{\textit{plib-runnable}}: Tasks that depend on third-party libraries that must be installed.
\item \textbf{\textit{class-runnable}}: Tasks with dependencies outside the function but within the same class.
\item \textbf{\textit{file-runnable}}: Tasks with dependencies outside the class but within the same source file.
\item \textbf{\textit{project-runnable}}: Tasks relying on code from other files within the same project.
\end{itemize}

Although CoderEval provides both Python and Java subsets, the Java subset suffers from an imbalanced distribution of dependency levels—for example, the \textit{class-runnable} and \textit{file-runnable} categories contain only 1 and 5 tasks, respectively. Such sparsity limits the reliability of reasoning behavior comparisons across categories. Moreover, a small-scale pilot study revealed no substantial differences in reasoning behaviors between Java and Python. Given these factors, and our primary interest in analyzing reasoning capabilities, we focus on the Python portion of the dataset.


Each task is presented to selected large reasoning models (LRMs) using the following prompt template that include both docstrings and function signatures. 
Across 230 tasks processed with the five LRMs, we obtain a total of 1,150 reasoning traces for analysis.


\begin{center}
\fbox{%
\begin{minipage}{0.9\linewidth}
You are a Python software engineer.
\newline
Generate Python code based on the following function signature and docstring. \newline
Output ONLY the code generated, in python markdown format. \newline
[Function Signature]\newline
    [Docstring]
\end{minipage}
}
\end{center}



\subsection{Manual Taxonomy Construction}

To analyze the reasoning processes and patterns exhibited by LRMs, we adopt an open coding protocol~\cite{opencoding} on the reasoning traces to manually construct a taxonomy of reasoning behaviors. The construction process consists of two phases: 

\textbf{Open-Coding Annotation.}  
We begin by randomly sampling tasks from each dependency level, ensuring statistical validity with a 95\% confidence level and a 5\% margin of error. This sampling yields 145 tasks and their corresponding 725 reasoning traces from five LRMs. For each reasoning trace, two annotators with more than five years programming experience independently examine the content to identify the underlying behaviors. They assign short, descriptive phrases, termed \textbf{\textit{reasoning actions}}, to sentences, paragraphs, or sections that represent fine-grained reasoning behaviors. During this process, the annotators use the original prompt (docstring and function signature) as context to accurately interpret the trace.  
Next, the annotators iteratively group similar codes into reasoning actions, refining the taxonomy through repeated review of both the codes and traces. A reasoning trace may be mapped to multiple categories when multiple reasoning actions are present. Disagreements between annotators are resolved through discussion with an arbitrator until full consensus is reached.

\textbf{Reliability Validation.}
To validate the reliability of the taxonomy, the two annotators independently annotate the remaining 425 reasoning traces (corresponding to 85 tasks) using the obtained coding schema. Each trace is reviewed to assign appropriate reasoning actions. If a trace cannot be classified, it is temporarily placed in a \textit{Pending} category. Inter-rater agreement is then measured using Cohen's Kappa~\cite{cohen1960coefficient}, yielding a value of 0.7054, which indicates substantial agreement~\cite{landis1977measurement}. This demonstrates the robustness of our coding schema and annotation process.  
Conflicts in annotation are resolved through discussion with the arbitrator. For reasoning traces classified as \textit{Pending}, the arbitrator also assists in identifying the underlying reasoning actions and determining whether new actions should be introduced. Through this process, we added a new reasoning action, \textit{Style Check}, and reassigned all \textit{Pending} traces into the updated taxonomy. The final taxonomy and annotations were reviewed and approved by all participants.

\vspace{-2mm}
\section{Reasoning Behavior Taxonomy}

\begin{figure*}
    \centering
    \includegraphics[width=\textwidth]{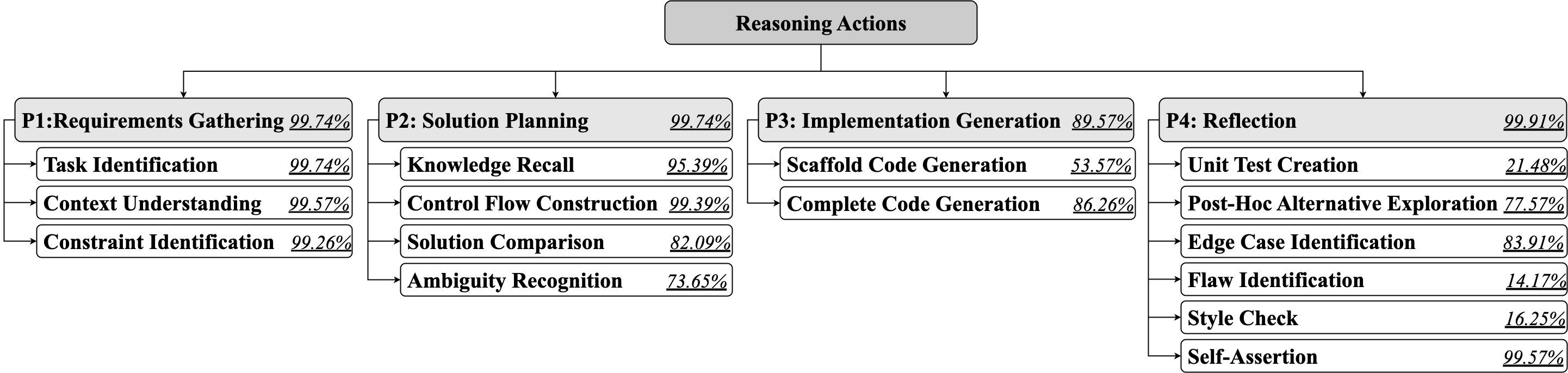}
    \vspace{-5mm}
    \caption{Reasoning action taxonomy in LRM-based code generation.}
    \label{fig:taxonomy}
\end{figure*}

Based on our manual annotations, we develop a taxonomy of 15 reasoning actions across 4 different phases, as illustrated in Figure~\ref{fig:taxonomy}. 

    
    
\noindent\textbf{Phase 1: Requirements Gathering}: In this phase, the LRMs collect and analyze all information necessary to understand the task specified in the user prompt. This involves identifying the task itself, interpreting the task-specific context, and detecting any constraints or conditions that might influence the solution. By performing these steps, the LRMs try to establish a clear understanding of both the functional intents and any additional requirements that will guide subsequent planning and code generation.
\begin{itemize}[leftmargin=15pt]
    \item \textbf{Task Identification (\TaskIdentification)}: At the start of the reasoning, the LRMs typically identify or reiterate the task description provided in the prompt.
    
    \example{... I need to generate Python code based on the given function signature and docstring. ...}{DeepSeek-R1-7B on Task 62b45df15108cfac7f2109dc}
    
    \item \textbf{Context Understanding (\ContextIdentification)}: The LRMs examine the task-specific context conveyed through the function signature and docstring in the user prompt, as these elements generally capture the functional intents. In our generation setting, this context primarily includes code construct elements such as variables, parameters, return values, and the basic functional logic described in the signature and docstring. For example, the case below shows DeepSeek-R1-7B recognizing code elements like \inlinecode{self} and \inlinecode{self.messages}, and analyzing the intended functionality of the generation. Moreover, we expect that supplying richer prompt context would lead to a broader set of analyzed contextual elements, such as retrieved functions in retrieval-augmented generation (RAG) settings.
     
    \example{... The function is called \inlinecode{status\_str} and it's an \ul{instance method} because it has \inlinecode{self} as the first parameter. The docstring says it should return a string by visiting the sorted \inlinecode{self.messages} list. For each element, it adds the prefix and the element. ...}{DeepSeek-R1-7B on Task 62b45df15108cfac7f2109dc}
    
    \item \textbf{Constraint Identification (\ConstraintIdentification)}: The LRMs identify limiting factors or constraints applied to inputs, outputs, or processing logic based on the given prompt. These constraints often represent \textit{additional} conditions that define either functional or non-functional requirements. For example, in the case below, the generated function is designed to be called recursively, which raises potential performance concerns that need to be addressed. This action is frequent because many docstrings and function signatures describe inputs or outputs in terms of specific value ranges or types.
    
    \example{... The docstring says that this function is called \ul{recursively} to update a partial \inlinecode{last\_applied\_manifest} from a \ul{Kubernetes API response}. ...}{DeepSeek-R1-7B on Task 62b869ebb4d922cb0e688cc6}
\end{itemize}

\noindent\textbf{Phase 2: Solution Planning}: In this phase, the LRMs rationalize about how to solve the task and begin structuring a solution before generating complete code. This involves recalling relevant programming knowledge (e.g., language features, libraries, or common idioms), constructing the control flow and data flow of the program, and, when necessary, comparing alternative approaches to determine the most suitable strategy. In addition, LRMs may recognize ambiguities or missing details in the prompt and prepare assumptions to resolve them in subsequent steps.
\begin{itemize}[leftmargin=15pt]
    \item \textbf{Knowledge Recall (\KnowledgeRecall)}: The LRMs recall or leverage prior knowledge about programming languages, libraries, data structures, or common coding patterns to guide subsequent solution steps. For example, in the case below, DeepSeek-R1-7B uses its knowledge of Python's \inlinecode{defaultdict} and its methods to determine the correct import and inform the next steps in the solution planning.

    \example{... Hmm, in Python, the \inlinecode{collections} module has a \inlinecode{defaultdict} that has a \inlinecode{popitem} method. That method removes and returns the least frequently used item. So maybe the code should import \inlinecode{defaultdict} from \inlinecode{collections}. ...}{DeepSeek-R1-7B on Task 62b8d23748ba5a41d1c3f497}
    
    \item \textbf{Control Flow Construction (\ControlFlowConstruction)}: The LRMs design the basic logical structure of the program, primarily including control flow and data flow, based on the task requirements and the context identified in the previous phase. This involves determining the sequence of operations, branching logic, and data handling steps. In the example below, DeepSeek-R1-7B constructs step-by-step logic, including module imports, method definition, and conditional checks, illustrating how it organizes control and data flow to solve the task.
    
    \example{... Then, define the \inlinecode{validate} method. Inside the method, I'll use \inlinecode{os.path.exists} to \ul{check if} the path is valid. \ul{If} it is, return \inlinecode{True}; \ul{else}, return \inlinecode{False}. ...}{DeepSeek-R1-7B on Task 62b45df05108cfac7f2109ce}
    
    \item \textbf{Solution Comparison (\ApproachComparison)}: After constructing a preliminary solution structure, the LRMs may compare alternative strategies. They evaluate different approaches in light of the current context and select the one most suitable for the user's request. In the example below, Qwen3-14B compares two similar functions, reasoning about input types and processing logic to determine an appropriate strategy. Note that, in our current setting without a full project context, the final choice among alternatives often relies on assumptions or random selection.
    
    \example{... \ul{Another approach:} look for similar functions. The other function is \inlinecode{update\_last\_appli-\\ed\_manifest\_dict\_from\_resp}, which \ul{probably} takes a dict and updates it with the response. So this function, for a list, would iterate over each item in the list and apply the dict function to each item. But the parameters here are lists, so \ul{maybe} the \inlinecode{observer\_schema} and response are lists of dicts, and the function processes each element in the list. ...}{Qwen3-14B on Task 62b869ebb4d922cb0e688cc6}
    
    \item \textbf{Ambiguity Recognition (\AmbiguityRecognition)}: The LRMs sometimes detect potential ambiguities or missing information in the prompt or current context. They flag these issues and prepare to resolve them by making ``reasonable'' assumptions in the subsequent phase. In the example below, Qwen3-14B identifies unclear instructions in the prompt, highlighting the need for assumptions or clarification. Such ambiguities often stem from vague docstrings or lack of awareness of the broader project context, including existing logic elsewhere in the codebase.

    \example{... The function signature has `commands' as a list, and `args' as another parameter. That's a bit unclear. \ul{Maybe} the `commands' is the main command, and `args' is the list of arguments. \ul{Or perhaps} `commands' is the list of commands to execute in sequence. But the docstring says "Run the given command(s) with the given arguments using a subprocess." \ul{So maybe it's supposed to run multiple commands, each with its own arguments?} ...}{Qwen3-8B on Task 62ece4992e6aefcf4aabbd83}

\end{itemize}

\noindent\textbf{Phase 3: Implementation Generation}: In this phase, the LRMs translate the solution plan designed in Phase 2 into concrete code implementation. The process typically occurs in stages, from scaffold code to a fully formed candidate function.

\begin{itemize}[leftmargin=15pt]

    \item \textbf{Scaffold Code Generation (\ScaffoldCodeGeneration)}: The LRMs sometimes generate partial or pseudo code that outlines the structure and main operations of the solution, without producing the final candidate code. This allows the models to roughly verify logic, identify missing pieces, or iteratively refine the implementation. In the example below, DeepSeek-R1-7B outputs a single expression for joining sorted messages with a prefix, illustrating the structure of the intended function without completing the full definition.

    \example{... Like this:\\
    \hspace*{1em}\inlinecode{return ", ".join(f"\{prefix\}\{msg\}" for msg in sorted(self.messages))} ...}{DeepSeek-R1-7B on Task 62b45df15108cfac7f2109dc}

    \item \textbf{Complete Code Generation (\CompleteCodeGeneration)}: After scaffolding, the LRMs may produce the complete candidate code that implements the solution. This output includes all necessary components, such as function definitions, control flow, and expressions, ready for review or execution. In the example below, DeepSeek-R1-7B combines the previously scaffolded expression into a full method definition, demonstrating how scaffolded ideas are integrated into a function.

    \example{... Putting it all together, the function becomes:\\
    \hspace*{1em}\inlinecode{def status\_str(self, prefix=""):}\\
    \hspace*{3em}\inlinecode{return ", ".join(f"\{prefix\}\{msg\}" for msg in sorted(self.messages))} ...}{DeepSeek-R1-7B on Task 62b45df15108cfac7f2109dc}

\end{itemize}

\noindent\textbf{Phase 4: Reflection}: In this phase, the LRMs critically evaluate the generated code to ensure it satisfies the task requirements, adheres to coding best practices, and handles possible edge cases. Reflection is both a verification and refinement phase, where the model reviews its own outputs, anticipates potential failures, and improves robustness and clarity. The main goals are correctness, reliability, maintainability, and alignment with the user's implicit or explicit expectations.
\begin{itemize}[leftmargin=15pt]
    \item \textbf{Unit Test Creation (\UnitTestCreation):} The LRMs generate test cases derived from the prompt, docstrings, or inferred requirements to explicitly verify code correctness. This ensures the generated code produces the expected outputs for typical inputs and helps uncover obvious bugs early. However, since the LRMs themselves perform the test execution and validation, the results may not fully guarantee reliability. In the example below, DeepSeek-R1-7B designs a test for a valid IP address to confirm that the function correctly identifies it as valid, illustrating how test cases are tied directly to the prompt requirements.
    
    \example{... Let me think about some test cases. For example: \inlinecode{"192.168.0.1"} should return \inlinecode{True}. ...}{DeepSeek-R1-7B on Task 62ece4992e6aefcf4aabbd84}
    
    \item \textbf{Post-Hoc Alternative Exploration (\PostHocAlternativeExploration):} The LRMs evaluate alternative strategies that differ from the initially generated code. This helps ensure that no potentially better approach is overlooked, and allows comparison against different libraries, algorithms, or coding patterns. In the example below, QWQ-32B considers multiple libraries for XML processing before settling on the correct \inlinecode{xmlsec} implementation, illustrating how the model iteratively explores alternatives for robustness. The difference from Solution Comparison (\ApproachComparison) is that \PostHocAlternativeExploration reflects whether alternative implementations are considered after generating code, whereas \ApproachComparison focuses on comparing potential solutions before committing to a concrete implementation.
    
    \example{... In reality, XML signatures can be more complex because they might sign specific parts of the document, use references, etc. \ul{Alternatively}, maybe the function is using the \inlinecode{xmlsec} library, which is a Python binding for XML Security. Oh right, \inlinecode{xmlsig} module might be part of that. ...}{QWQ-32B on Task 630629d052e177c0ba46a0a1}
    
    \item \textbf{Edge Case Identification (\EdgeCaseIdentification):} The LRMs proactively reason about uncommon or extreme input scenarios that could trigger failures. This aims to strengthen code robustness by ensuring proper handling of unexpected or incomplete data. In the example below, Qwen3-1.7B identifies missing dictionary keys and empty lists, demonstrating attention to possible input anomalies and safe handling of these cases.
    
    \example{... Now, considering possible edge cases: what if the fixity is empty? Or if the \inlinecode{`files'} key is missing? ...}{Qwen3-1.7B on Task 62b45e175108cfac7f210a19}
    
    \item \textbf{Flaw Identification (\FlawIdentification):} The LRMs examine the code for logical or semantic errors and revise it to correct mistakes. This step ensures that the solution behaves as intended across all valid inputs. In the example below, QWQ-32B identifies that using \inlinecode{None} would incorrectly filter triples and revises the logic to account for all object nodes, illustrating careful reasoning about functional correctness.
    
    \example{... But wait, the triple is (node, prop, parent). So, the object of the triple is the parent. So, if there's any such triple, then the node has a parent. In the code above, the triple is (node, prop, None), which would return all triples where the subject is node, the predicate is prop, and the object is None. \ul{But that's not correct. Because the object can be any node.} ...}{QWQ-32B on Task 630629d052e177c0ba46a0a1}
    
    \item \textbf{Style Check (\StyleCheck):} The LRMs evaluate whether the code adheres to coding conventions, formatting standards, and documentation best practices. Proper style ensures readability, maintainability, and consistency with user expectations. In the example below, DeepSeek-R1-7B confirms that the code is properly indented, syntactically correct, and well-documented, highlighting the model’s attention to presentation and clarity.
    
    \example{... I'll make sure the code is properly indented and follows Python syntax. ...}{DeepSeek-R1-7B Task ID: 62b8d22f48ba5a41d1c3f488} 
    
    \item \textbf{Self-Assertion (\SelfAssertion):} The LRMs provide final closure by asserting that the generated code fulfills the task requirements. This step signals confidence in the correctness and completeness of the solution. In the example below, Qwen3-8B reaffirms that the code correctly initializes the \inlinecode{\_Converter} instance and satisfies the intended function, demonstrating the model's self-assertion process.
    
    \example{... I think that's \ul{a reasonable approach}. So the final code would be as above. But since the user hasn't provided specifics, this is an \ul{assumption}. However, given the problem statement, this seems like \ul{a logical implementation}. ...}{Qwen3-8B on Task 62b45df05108cfac7f2109ce}
\end{itemize}

\finding{We identified 15 reasoning actions across four phases: Requirements Gathering, Solution Planning, Implementation Generation, and Reflection. Nearly all reasoning traces include all four phases, with only a small portion (10\%) omitting Implementation Generation. Among the 15 actions, most appear frequently in the reasoning process, while Scaffold Code Generation (\ScaffoldCodeGeneration), Unit Test Creation (\UnitTestCreation), Flaw Identification (\FlawIdentification), and Style Check (\StyleCheck) occur relatively less often ($< 50\%$).}

\section{Empirical Results}
Based on the taxonomy, we conduct a series of experiments and present empirical results to address the research questions.

\subsection{RQ1: Common Reasoning Patterns}\label{sec:patterns}
We count the reasoning actions in the traces and analyze the common patterns exhibited by the LRMs. A reasoning pattern is defined as the complete combination of individual reasoning actions that appear within a single trace.

\textbf{Overall Results.} Table \ref{tab:freq_pattern} presents the top five most common reasoning patterns observed in reasoning traces. The most frequent pattern, FP1, accounts for approximately 17\% of all traces. This aligns with the action frequencies, as all actions within FP1 appear in more than half ($>50\%$) of the collected traces. This frequent pattern generally reflects a human-like coding process, encompassing requirement analysis from multiple perspectives, clarification of ambiguities, comparison of alternative solutions, code implementation, and subsequent review to identify potential defects.
Frequent patterns FP3, FP4, and FP5 are extensions of FP1, incorporating \UnitTestCreation, \FlawIdentification, or \StyleCheck, respectively. These variations typically depend on the specific challenges faced by the LRMs: for example, when validation or verification is required, the model performs Unit Test Creation (\UnitTestCreation) to clarify logic; when inconsistencies are noticed, it performs Flaw Identification (\FlawIdentification); and when readability or conventions are at stake, it enforces code Style Checks (\StyleCheck). {While FP2 is a simplified version of FP1, where the LRM perceived that it is not necessary to perform the intermediary step of Scaffold Code Generation (\ScaffoldCodeGeneration) and directly work on the final code.}


\begin{table*}
\centering
\footnotesize
\caption{Top-5 common reasoning action patterns across all reasoning traces. The most frequent pattern, along with its occurrences in other patterns, is \hlyellow{highlighted}.}
\label{tab:freq_pattern}
\vspace{-3mm}
\begin{tabular}{l|r}
\toprule
\textbf{Action Pattern} & \textbf{Frequency \# (\%)} \\
\midrule
\textbf{FP1:} \hlyellow{\TaskIdentification, \ContextIdentification, \ConstraintIdentification, \KnowledgeRecall, \ControlFlowConstruction, \ApproachComparison, \AmbiguityRecognition, \ScaffoldCodeGeneration, \CompleteCodeGeneration, \PostHocAlternativeExploration, \EdgeCaseIdentification, \SelfAssertion} & 207 (17.48\%) \\
\textbf{FP2:} \TaskIdentification, \ContextIdentification, \ConstraintIdentification, \KnowledgeRecall, \ControlFlowConstruction, \ApproachComparison, \AmbiguityRecognition, \CompleteCodeGeneration, \PostHocAlternativeExploration, \EdgeCaseIdentification, \SelfAssertion & 188 (16.35\%) \\
\textbf{FP3:} \hlyellow{\TaskIdentification, \ContextIdentification, \ConstraintIdentification, \KnowledgeRecall, \ControlFlowConstruction, \ApproachComparison, \AmbiguityRecognition, \ScaffoldCodeGeneration, \CompleteCodeGeneration}, \UnitTestCreation, \hlyellow{\PostHocAlternativeExploration, \EdgeCaseIdentification, \SelfAssertion} & 57 (4.96\%) \\
\textbf{FP4:} \hlyellow{\TaskIdentification, \ContextIdentification, \ConstraintIdentification, \KnowledgeRecall, \ControlFlowConstruction, \ApproachComparison, \AmbiguityRecognition, \ScaffoldCodeGeneration, \CompleteCodeGeneration, \PostHocAlternativeExploration, \EdgeCaseIdentification}, \FlawIdentification, \hlyellow{\SelfAssertion} & 53 (5.61\%) \\
\textbf{FP5:} \hlyellow{\TaskIdentification, \ContextIdentification, \ConstraintIdentification, \KnowledgeRecall, \ControlFlowConstruction, \ApproachComparison, \AmbiguityRecognition, \ScaffoldCodeGeneration, \CompleteCodeGeneration, \PostHocAlternativeExploration, \EdgeCaseIdentification,} \StyleCheck, \hlyellow{\SelfAssertion} & 52 (4.52\%) \\\hline
\textbf{Others} & 599 (52.09\%)\\
\bottomrule
\end{tabular}
\end{table*}

The notable outlier among the top five is FP5. Unlike the others, FP5 omits Implementation Generation and Reflection. In this pattern, after performing Requirements Gathering (P1) and Solution Planning (P2), the LRMs directly output generated code as the final answer without additional reasoning or review on implementation. This behavior tends to occur in straightforward tasks where the solution is clear and no further analysis is deemed necessary.


\begin{figure}
    \vspace{-3mm}
    \centering
    \footnotesize
    \fontfamily{ptm}\selectfont
    \begin{tikzpicture}
    \pie[
      text=legend,
      sum=auto,
      radius=2,
      before number=\scriptsize,
      after number=\%,
      color={color1, color2, color3, color4, color5, color6}
    ]{
      8.53/\textit{self-contained},
      6.72/\textit{slib-runnable},
      6.53/\textit{plib-runnable},
      30.85/\textit{class-runnable},
      37.57/\textit{file-runnable},
      9.8/\textit{project-runnable}
    }
  \end{tikzpicture}
    \caption{Distribution of the top 5 common patterns across different dependency levels.}
    \label{fig:top5-scope}
    \vspace{-3mm}
\end{figure}
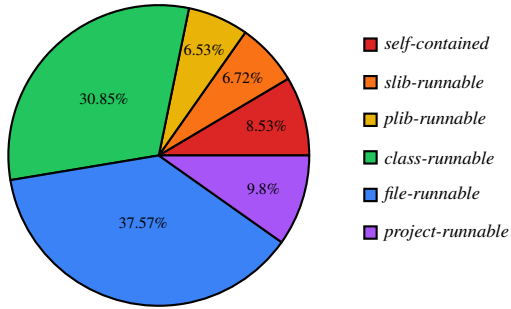

\textbf{Breakdown Results.} Figure \ref{fig:top5-scope} shows how the top five patterns are distributed across different dependency levels. These patterns appear most often in complex levels such as \textit{class-runnable}, \textit{file-runnable}, and \textit{project-runnable}, where solving the tasks require a broader context. This suggests that LRMs tend to fall back on recurring reasoning routines when dealing with complex or underspecified problems. While providing stability, this also introduces extra reasoning steps and overhead for simpler tasks at the \textit{self-contained}, \textit{slib-runnable}, and \textit{plib-runnable} levels. For those more straightforward cases, LRMs usually apply simpler patterns without Ambiguity Recognition (\AmbiguityRecognition), since the prompts often already define problems clearly without relying on in-project context.

\finding{LRMs most often follow a human-like coding workflow of analyzing requirements, clarifying ambiguities, comparing solutions, implementing code, and reviewing for defects, with variations such as scaffolding, flaw detection, or style checks depending on task difficulty. These patterns appear most frequently in complex dependency levels, while simpler tasks are handled with lighter reasoning that avoids unnecessary steps like ambiguity recognition.}

\subsection{RQ2: Comparison Across Models}



Table~\ref{tab:LRM-pattern-combination} presents the top 5 most common reasoning patterns for each LRM. The results indicate that the Qwen3 models exhibit nearly identical dominant patterns, suggesting that models within the same series display highly consistent reasoning behaviors regardless of parameter size. Notably, the highlighted pattern corresponds to FP1 introduced in Section~\ref{sec:patterns}, which explains its strong prevalence across reasoning traces. Although QWQ-32B places slightly greater emphasis on Style Check (\StyleCheck), its overall reasoning patterns remain broadly consistent with the Qwen3 models, reflecting its role as their predecessor developed by the same group. By contrast, DeepSeek-R1-7B often deviates from this trend, frequently omitting the Implementation Generation phase in its reasoning trace. Moreover, in our experiments, DeepSeek-R1-7B follows a more \textit{linear, waterfall-like} reasoning style, whereas other models adopt a more \textit{cyclical, iterative} approach. This difference is evident in its lower frequency of Solution Comparison (\ApproachComparison), Ambiguity Recognition (\AmbiguityRecognition), Edge Case Identification (\EdgeCaseIdentification), and Alternative Comparison (\PostHocAlternativeExploration), which occur in only 15–46\% of traces compared to over 80\% in the Qwen3 and QWQ-32B models' traces. Such divergence underscores the strong influence of training data, as Qwen-family models and DeepSeek models were trained on different reasoning traces, which in turn shaped their distinct reasoning styles.

\begin{table*}
\caption{Top 5 common reasoning patterns across LRMs. The most frequent pattern (FP1 in RQ1) shared among LRMs is \hlyellow{highlighted}.}
\label{tab:LRM-pattern-combination}
\centering
\vspace{-3mm}
\resizebox{\textwidth}{!}{%
\begin{tabular}{p{4cm} p{4cm} p{4cm} p{4cm} p{4cm}}
\toprule
\textbf{DeepSeek-R1-7B} & \textbf{Qwen3-1.7B} & \textbf{Qwen3-8B} & \textbf{Qwen3-14B} & \textbf{QWQ-32B} \\
\midrule
\TaskIdentification, \ContextIdentification, \ConstraintIdentification, \KnowledgeRecall, \ControlFlowConstruction, \SelfAssertion
  & \TaskIdentification, \ContextIdentification, \ConstraintIdentification, \KnowledgeRecall, \ControlFlowConstruction, \ApproachComparison, \AmbiguityRecognition, \CompleteCodeGeneration, \PostHocAlternativeExploration, \EdgeCaseIdentification, \SelfAssertion
  & \hlyellow{\TaskIdentification, \ContextIdentification, \ConstraintIdentification, \KnowledgeRecall, \ControlFlowConstruction, \ApproachComparison, \AmbiguityRecognition, \ScaffoldCodeGeneration, \CompleteCodeGeneration, \PostHocAlternativeExploration, \EdgeCaseIdentification, \SelfAssertion}
  & \hlyellow{\TaskIdentification, \ContextIdentification, \ConstraintIdentification, \KnowledgeRecall, \ControlFlowConstruction, \ApproachComparison, \AmbiguityRecognition, \ScaffoldCodeGeneration, \CompleteCodeGeneration, \PostHocAlternativeExploration, \EdgeCaseIdentification, \SelfAssertion}
  & \hlyellow{\TaskIdentification, \ContextIdentification, \ConstraintIdentification, \KnowledgeRecall, \ControlFlowConstruction, \ApproachComparison, \AmbiguityRecognition, \ScaffoldCodeGeneration, \CompleteCodeGeneration, \PostHocAlternativeExploration, \EdgeCaseIdentification, \SelfAssertion} \\\hline
\TaskIdentification, \ContextIdentification, \ConstraintIdentification, \KnowledgeRecall, \ControlFlowConstruction, \StyleCheck, \SelfAssertion
  & \hlyellow{\TaskIdentification, \ContextIdentification, \ConstraintIdentification, \KnowledgeRecall, \ControlFlowConstruction, \ApproachComparison, \AmbiguityRecognition, \ScaffoldCodeGeneration, \CompleteCodeGeneration, \PostHocAlternativeExploration, \EdgeCaseIdentification, \SelfAssertion}
  & \TaskIdentification, \ContextIdentification, \ConstraintIdentification, \KnowledgeRecall, \ControlFlowConstruction, \ApproachComparison, \AmbiguityRecognition, \CompleteCodeGeneration, \PostHocAlternativeExploration, \EdgeCaseIdentification, \SelfAssertion
  & \TaskIdentification, \ContextIdentification, \ConstraintIdentification, \KnowledgeRecall, \ControlFlowConstruction, \ApproachComparison, \AmbiguityRecognition, \CompleteCodeGeneration, \PostHocAlternativeExploration, \EdgeCaseIdentification, \SelfAssertion
  & \TaskIdentification, \ContextIdentification, \ConstraintIdentification, \KnowledgeRecall, \ControlFlowConstruction, \ApproachComparison, \AmbiguityRecognition, \ScaffoldCodeGeneration, \CompleteCodeGeneration, \PostHocAlternativeExploration, \EdgeCaseIdentification, \StyleCheck, \SelfAssertion \\\hline
\hlyellow{\TaskIdentification, \ContextIdentification, \ConstraintIdentification, \KnowledgeRecall, \ControlFlowConstruction, \ApproachComparison, \AmbiguityRecognition, \ScaffoldCodeGeneration, \CompleteCodeGeneration, \PostHocAlternativeExploration, \EdgeCaseIdentification, \SelfAssertion}
  & \TaskIdentification, \ContextIdentification, \ConstraintIdentification, \KnowledgeRecall, \ControlFlowConstruction, \ApproachComparison, \AmbiguityRecognition, \ScaffoldCodeGeneration, \CompleteCodeGeneration, \PostHocAlternativeExploration, \EdgeCaseIdentification, \FlawIdentification, \SelfAssertion
  & \TaskIdentification, \ContextIdentification, \ConstraintIdentification, \KnowledgeRecall, \ControlFlowConstruction, \ApproachComparison, \AmbiguityRecognition, \ScaffoldCodeGeneration, \CompleteCodeGeneration, \PostHocAlternativeExploration, \EdgeCaseIdentification, \FlawIdentification, \SelfAssertion
  & \TaskIdentification, \ContextIdentification, \ConstraintIdentification, \KnowledgeRecall, \ControlFlowConstruction, \ApproachComparison, \AmbiguityRecognition, \ScaffoldCodeGeneration, \CompleteCodeGeneration, \PostHocAlternativeExploration, \EdgeCaseIdentification, \FlawIdentification, \SelfAssertion
  & \TaskIdentification, \ContextIdentification, \ConstraintIdentification, \KnowledgeRecall, \ControlFlowConstruction, \ApproachComparison, \AmbiguityRecognition, \ScaffoldCodeGeneration, \CompleteCodeGeneration, \UnitTestCreation, \PostHocAlternativeExploration, \EdgeCaseIdentification, \SelfAssertion \\\hline
\TaskIdentification, \ContextIdentification, \ConstraintIdentification, \KnowledgeRecall, \ControlFlowConstruction, \ApproachComparison, \SelfAssertion
  & \TaskIdentification, \ContextIdentification, \ConstraintIdentification, \KnowledgeRecall, \ControlFlowConstruction, \ApproachComparison, \AmbiguityRecognition, \CompleteCodeGeneration, \PostHocAlternativeExploration, \EdgeCaseIdentification, \StyleCheck, \SelfAssertion
  & \TaskIdentification, \ContextIdentification, \ConstraintIdentification, \KnowledgeRecall, \ControlFlowConstruction, \ApproachComparison, \AmbiguityRecognition, \ScaffoldCodeGeneration, \CompleteCodeGeneration, \PostHocAlternativeExploration, \EdgeCaseIdentification, \StyleCheck, \SelfAssertion
  & \TaskIdentification, \ContextIdentification, \ConstraintIdentification, \KnowledgeRecall, \ControlFlowConstruction, \ApproachComparison, \AmbiguityRecognition, \ScaffoldCodeGeneration, \CompleteCodeGeneration, \UnitTestCreation, \PostHocAlternativeExploration, \EdgeCaseIdentification, \SelfAssertion
  & \TaskIdentification, \ContextIdentification, \ConstraintIdentification, \KnowledgeRecall, \ControlFlowConstruction, \ApproachComparison, \AmbiguityRecognition, \CompleteCodeGeneration, \PostHocAlternativeExploration, \EdgeCaseIdentification, \SelfAssertion \\\hline
\TaskIdentification, \ContextIdentification, \ConstraintIdentification, \KnowledgeRecall, \ControlFlowConstruction, \ScaffoldCodeGeneration, \SelfAssertion
  & \TaskIdentification, \ContextIdentification, \ConstraintIdentification, \KnowledgeRecall, \ControlFlowConstruction, \ApproachComparison, \AmbiguityRecognition, \ScaffoldCodeGeneration, \CompleteCodeGeneration, \PostHocAlternativeExploration, \EdgeCaseIdentification, \StyleCheck, \SelfAssertion
  & \TaskIdentification, \ContextIdentification, \ConstraintIdentification, \KnowledgeRecall, \ControlFlowConstruction, \ApproachComparison, \AmbiguityRecognition, \CompleteCodeGeneration, \PostHocAlternativeExploration, \EdgeCaseIdentification, \FlawIdentification, \SelfAssertion
  & \TaskIdentification, \ContextIdentification, \ConstraintIdentification, \KnowledgeRecall, \ControlFlowConstruction, \ApproachComparison, \AmbiguityRecognition, \CompleteCodeGeneration, \EdgeCaseIdentification, \SelfAssertion
  & \TaskIdentification, \ContextIdentification, \ConstraintIdentification, \KnowledgeRecall, \ControlFlowConstruction, \ApproachComparison, \AmbiguityRecognition, \ScaffoldCodeGeneration, \CompleteCodeGeneration, \UnitTestCreation, \PostHocAlternativeExploration, \EdgeCaseIdentification, \StyleCheck, \SelfAssertion \\
\bottomrule
\end{tabular}%
}
\end{table*}

\finding{LRMs in the Qwen3 series exhibit highly similar reasoning behaviors across parameter sizes, generally following an \textit{iterative, cyclical} style. In contrast, DeepSeek-R1-7B adopts a more \textit{linear, waterfall-like} reasoning pattern, reflecting a distinct approach that may stem from differences in the reasoning traces used during training.}

\subsection{RQ3: Impact on Correctness}
\label{sub:correctness}
To assess the impact of reasoning behaviors on functional correctness, we analyze the statistical correlation between the Pass@1 metric and the observed reasoning actions and patterns. Table~\ref{tab:pass-rate} reports Pass@1 results for each LRM across different dependency levels. The results for Qwen3 models with and without reasoning indicate that enabling reasoning has minimal impact on correctness. Consequently, our analysis focuses on the characteristics of reasoning behaviors rather than the binary setting of reasoning enabled versus disabled. For DeepSeek-R1-7B and QWQ-32B, which do not support disabling reasoning, only results with reasoning (\textit{w/ R}) are reported.

\begin{table*}
    \centering
    \footnotesize
    \caption{Pass@1 scores on code generation tasks (\%). \textit{w/ R} and \textit{w/o R} denote results with and without reasoning enabled, respectively. Since DeepSeek-R1-7B and QWQ-32B do not support disabling reasoning, only \textit{w/ R} results are reported for these models. }
    \label{tab:pass-rate}
    \begin{tabular}{l|c cc cc cc c}
    \toprule
    \multirow{2}{*}{\textbf{Category}} & \textbf{DeepSeek-R1-7B} & \multicolumn{2}{c}{\textbf{Qwen3-1.7B}} & \multicolumn{2}{c}{\textbf{Qwen3-8B}} & \multicolumn{2}{c}{\textbf{Qwen3-14B}} & \textbf{QWQ-32B} \\
    \cmidrule(r){2-2}\cmidrule(lr){3-4} \cmidrule(lr){5-6} \cmidrule(lr){7-8}\cmidrule(r){9-9}
    & \textit{w/ R} & \textit{w/ R} & \textit{w/o R} & \textit{w/ R} & \textit{w/o R} & \textit{w/ R} & \textit{w/o R} & \textit{w/ R} \\
    \midrule
    \textit{self-contained}    & 28.57 & 40.00 & 42.86 & 51.43 & 48.57 & 51.43 & 48.57 & 54.29 \\
    \textit{slib-runnable}     & 25.00 & 39.29 & 35.71 & 50.00 & 50.00 & 50.00 & 53.57 & 46.43 \\
    \textit{plib-runnable}     & 19.05  & 19.05 & 23.81  & 28.57 & 28.57  & 28.57 & 28.57  & 38.10  \\
    \textit{class-runnable}    & 10.91 & 12.73 & 14.55 & 23.64 & 14.55 & 23.64 & 21.82 & 18.52 \\
    \textit{file-runnable}     & 10.29  & 17.65 & 23.53  & 25.00 & 23.53  & 25.00 & 26.47  & 19.12  \\
    \textit{project-runnable}  & 4.35   & 4.35  & 13.04 & 4.35  & 8.70 & 13.04 & 4.35 & 4.35   \\
    \midrule
    \textbf{Overall}  & 15.22 & 21.30 & 24.78 & 29.57 & 27.39 & 30.87 & 30.00 & 27.83 \\
    \bottomrule
    \end{tabular}
\end{table*}

\vspace{-2mm}
\subsubsection{Individual Reasoning Actions}
We employ the phi-coefficient ($\varphi$)~\cite{cramer1999mathematical} to quantify the correlation between each reasoning action and the correctness of the generated code. A positive $\varphi$ indicates that the presence of an action correlates with higher correctness, a negative $\varphi$ suggests the action is associated with lower correctness, and values near zero imply little to no correlation.

\begin{figure*}
    \centering
    \includegraphics[width=0.7\linewidth]{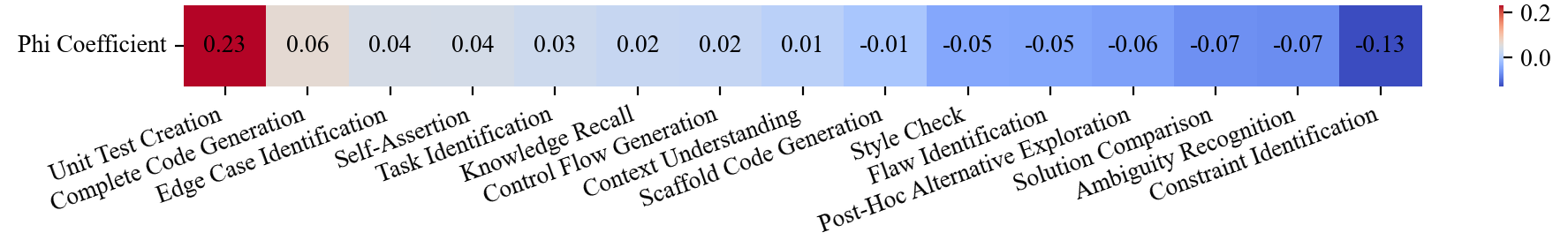}
    \vspace{-5mm}
    \caption{Correlation between reasoning actions and code correctness.}
    \label{fig:phi-heatmap}
\end{figure*}

Figure~\ref{fig:phi-heatmap} presents the correlations between each reasoning action and the correctness of the generated code. Among all actions, Unit Test Creation (\UnitTestCreation) shows the strongest influence, with a weak positive correlation to correctness. This indicates that generating unit tests helps LRMs validate their own logic, promoting more thorough reasoning and increasing the chance of producing correct outputs.
By contrast, Constraint Identification (\ConstraintIdentification), Ambiguity Recognition (\AmbiguityRecognition), and Solution Comparison (\ApproachComparison) exhibit weak negative correlations. These actions often appear when prompts and contexts are unclear or under-specified, leading LRMs to make additional assumptions. Such assumptions may be incorrect, while overly rigid constraints can prematurely narrow the solution space, and repeated ambiguity recognition or comparison may lead to reasoning loops and inefficiency. As a result, these behaviors increase cognitive load without reliably improving—and sometimes even reducing—the correctness of the final solution.

\vspace{-2mm}
\subsubsection{Combined Reasoning Actions}
We further examine whether specific combinations of reasoning actions influence the correctness of the generated code. To this end, we apply the Apriori algorithm~\cite{huang2000fast}, a classic association rule mining method commonly used to discover frequent itemsets and their correlations within large datasets. By treating reasoning actions as items and reasoning traces as transactions, Apriori allows us to identify combinations of actions that are strongly associated with successful code generation outcomes. It is worth noting that the mined combinations may be subsequences of the complete reasoning patterns identified in RQ1.

\begin{table}
\centering
\footnotesize
\vspace{-3mm}
\caption{Top 10 mined common action combinations leading to passing tests.}
\label{tab:assoc_rules}
\vspace{-4mm}
\begin{tabular}{c|l}
\toprule
\textbf{No} & \textbf{Action Combinations} \\
\midrule
1 & \TaskIdentification, \UnitTestCreation, \SelfAssertion\\
2 & \TaskIdentification, \UnitTestCreation\\
3 & \TaskIdentification, \ContextIdentification, \UnitTestCreation, \SelfAssertion\\
4 & \TaskIdentification, \ControlFlowConstruction, \UnitTestCreation, \SelfAssertion\\
5 & \TaskIdentification, \ContextIdentification, \UnitTestCreation\\
6 & \TaskIdentification, \ControlFlowConstruction, \UnitTestCreation\\
7 & \TaskIdentification, \ContextIdentification, \ControlFlowConstruction, \UnitTestCreation, \SelfAssertion\\
8 & \TaskIdentification, \ContextIdentification, \ControlFlowConstruction, \UnitTestCreation\\
9 & \TaskIdentification, \UnitTestCreation, \EdgeCaseIdentification, \SelfAssertion\\
10 & \TaskIdentification, \KnowledgeRecall, \UnitTestCreation, \SelfAssertion\\
\bottomrule
\end{tabular}
\vspace{-5mm}
\end{table}

\textbf{Overall Results.} Using this approach, we mine several action combinations that are linked to generated code successfully passing all test cases for a given task. Table \ref{tab:assoc_rules} reports the top 10 such combinations, where the ratio indicates the proportion of conversions resulting in correct solutions. The presence and co-occurrence of Unit Test Creation (\UnitTestCreation), Self-Assertion (\SelfAssertion), and Task Identification (\TaskIdentification) are positively correlated with higher success rates. Consistent with the findings in the previous section, \UnitTestCreation emerges as the strongest individual predictor of correctness, while its combination with \TaskIdentification and \SelfAssertion achieves the highest conversion ratio from generation attempts to correct outputs. This suggests that by having a deeper understanding of the problem, LRMs can construct more comprehensive unit tests, which improves their ability to self-assert solutions and increases the likelihood of passing test cases.

\textbf{Breakdown Results.} Breaking down the results by task dependency level, Table \ref{tab:pattern-dependency-pass} presents the top-10 combinations mined by the Apriori algorithm that lead to passing all test cases. Due to the sparsity of pass results on higher complexity level tasks, no common combinations emerge even with a minimal confidence threshold (0.02) at the \textit{project-runnable} level tasks, alongside a significant drop in confidence level for \textit{class-runnable} and \textit{file-runnable} tasks of around $0.2 -0.3$ compared to the lower complexity tasks with a confidence value of $>0.5$. Consistent with the overall results, the overlaps across the five levels highlight Unit Test Creation (\UnitTestCreation) as the most frequent action associated with correct outcomes. 

\begin{table*}
\caption{Top 10 mined combinations across dependency types leading to test pass. Pass results are sparse at the project-runnable levels, preventing the emergence of common combinations even under a minimal confidence threshold (0.02). Highlighted are the frequent intersections shared across all combinations.}
\label{tab:pattern-dependency-pass}
\centering
\vspace{-3mm}
\resizebox{\textwidth}{!}{%
\begin{tabular}{c|p{4cm} p{4cm} p{4cm} p{4cm} p{4cm}}
\toprule
\textbf{No} & \textbf{\textit{self-contained}} & \textbf{\textit{slib-runnable}} & \textbf{\textit{plib-runnabl}e} & \textbf{\textit{class-runnable}} & \textbf{\textit{file-runnable}}  \\
\midrule
1 & \TaskIdentification, \UnitTestCreation, \SelfAssertion & \TaskIdentification, \ContextIdentification, \ControlFlowConstruction, \UnitTestCreation, \SelfAssertion &TSK, CTX, CST, KRL, CFL, UTC, SFA & TSK, CTX, CST, KRL, CFL, SFA & TSK, CTX, CST, KRL, CFL, SFA\\
2 & \TaskIdentification, \ContextIdentification, \UnitTestCreation, \SelfAssertion & \TaskIdentification, \ContextIdentification, \ControlFlowConstruction, \UnitTestCreation &TSK, CTX, CST, KRL, CFL, UTC, EGC, SFA & TSK, CTX, CST, KRL, SFA & TSK, CTX, CST, KRL, SFA\\
3 &\TaskIdentification, \ControlFlowConstruction, \UnitTestCreation, \SelfAssertion & \TaskIdentification, \ContextIdentification, \KnowledgeRecall, \ControlFlowConstruction, \UnitTestCreation, \SelfAssertion &TSK, CTX, CST, KRL, CFL, CCG, UTC, SFA & TSK, CTX, KRL, CFL, SFA & TSK, CTX, CST, CFL, SFA\\
4 &\TaskIdentification, \ContextIdentification, \ControlFlowConstruction, \UnitTestCreation, \SelfAssertion & \TaskIdentification, \ContextIdentification, \KnowledgeRecall, \ControlFlowConstruction, \UnitTestCreation  &TSK, CTX, CST, KRL, CFL, CCG, UTC, EGC, SFA & TSK, CTX, KRL, SFA & TSK, CTX, CST, SFA\\
5 &\TaskIdentification, \UnitTestCreation, \EdgeCaseIdentification, \SelfAssertion & \TaskIdentification, \ContextIdentification, \ControlFlowConstruction, \CompleteCodeGeneration, \UnitTestCreation, \SelfAssertion &TSK, CTX, CST, KRL, CFL, UTC, ALT, EGC, SFA & TSK, CTX, CST, CFL, SFA & TSK, CTX, CST, KRL, CFL, EGC, SFA\\
6 &\TaskIdentification, \ContextIdentification, \UnitTestCreation, \EdgeCaseIdentification, \SelfAssertion & \TaskIdentification, \ContextIdentification, \ControlFlowConstruction, \CompleteCodeGeneration, \UnitTestCreation &TSK, CTX, CST, KRL, CFL, UTC, ALT, SFA & TSK, CTX, CST, CFL & TSK, CTX, CST, KRL, CFL, CCG, SFA\\
7 &\TaskIdentification, \ControlFlowConstruction, \UnitTestCreation, \EdgeCaseIdentification, \SelfAssertion & \TaskIdentification, \ContextIdentification, \ControlFlowConstruction, \UnitTestCreation, \EdgeCaseIdentification, \SelfAssertion &TSK, CTX, CST, KRL, CFL, CCG, UTC, ALT, EGC, SFA & TSK, CTX, CST, SFA & TSK, CTX, CST, CFL, EGC, SFA\\
8 &\TaskIdentification, \ContextIdentification, \ControlFlowConstruction, \UnitTestCreation, \EdgeCaseIdentification, \SelfAssertion & \TaskIdentification, \ContextIdentification, \ControlFlowConstruction, \UnitTestCreation, \EdgeCaseIdentification &TSK, CTX, CST, KRL, CFL, CCG, UTC, ALT, SFA & TSK, CTX, CST & TSK, CTX, CST, CFL, CCG, SFA\\
9 &\TaskIdentification, \ContextIdentification, \KnowledgeRecall, \UnitTestCreation, \SelfAssertion & \TaskIdentification, \ContextIdentification, \KnowledgeRecall, \ControlFlowConstruction, \UnitTestCreation, \EdgeCaseIdentification &TSK, CTX, CST, KRL, CFL, SCG, CCG, UTC, ALT, EGC, SFA & TSK, CTX, CFL, SFA & TSK, CTX, CST, KRL, CFL, CCG, EGC, SFA\\
10 &\TaskIdentification, \ContextIdentification, \CompleteCodeGeneration, \UnitTestCreation, \SelfAssertion & \TaskIdentification, \ContextIdentification, \KnowledgeRecall, \ControlFlowConstruction, \UnitTestCreation, \EdgeCaseIdentification &TSK, CTX, CST, KRL, CFL, SCG, UTC, SFA & TSK, CTX, CFL & TSK, KRL, CFL, CCG, EGC, SFA \\
\bottomrule
\end{tabular}%
}
\vspace{-3mm}
\end{table*}

Beyond these shared actions, each dependency level exhibits distinct patterns. There are noticeably fewer diverse actions for \textit{self-contained} tasks. This reflects that such tasks with no dependencies and reduced complexity allow LRMs to focus on identifying tasks, and perform verification through Unit Test Creation (\UnitTestCreation). \textit{Slib-runnable} tasks, which rely on standard libraries, are typically functionally well-defined and thus elicit relatively straightforward reasoning behaviours. With the reliance in standard libraries, LRM would need to perform Context Identification (\ContextIdentification) to distinguish the need for third-party or standard libraries and perform the least Self-Assertion (\SelfAssertion), which could indicate that LRM is overly confident in the generated code. In contrast, \textit{plib-runnable} tasks frequently involve Knowledge Recall (\KnowledgeRecall) and Alternative Exploration (\PostHocAlternativeExploration) with both implementation actions of Scaffold Code Generation (\ScaffoldCodeGeneration) and Complete Code Generation (\CompleteCodeGeneration), indicating that LRMs tend to explore key components, such as usage patterns of public library APIs and verify the results, compare with solutions using other libraries, before producing a final, correct implementation. \textit{Class-runnable} tasks, which rely on code defined within the same class, have a noticeable lack of Edge Case Generation (EGC), Unit Test Creation (\UnitTestCreation), and Complete Code Generation in (\CompleteCodeGeneration), suggesting that LRMs prioritize compatibility with defined code over robustness and exploratory code generation. Finally, \textit{file-runnable} tasks with dependencies outside of the class require a lot of context not provided within the docstrings, and the LRM also performed fewer Unit Test Creation (\UnitTestCreation) relative to tasks that require lower-level dependencies. However, unlike \textit{class-runnable} tasks, \textit{file-runnable} tasks LRMs would perform Complete Code Generation (\CompleteCodeGeneration) and Edge Case Generation (\EdgeCaseIdentification), suggesting that at this dependency level, LRMs may be less able to produce final code directly and instead perform additional reasoning processes to ensure correctness of the output.
The differences across dependency levels demonstrate that LRMs can, to some extent, adapt their reasoning behaviours to produce accurate solutions based on the provided task descriptions (docstrings) and code contexts (function signatures).

\finding{LRMs adapt their reasoning strategies based on task dependency levels: simpler, self-contained or standard-library tasks trigger minimal, focused reasoning with unit test creation, while tasks relying on public libraries or external files induce more exploratory and verification-focused behaviors. Tasks with intra-class dependencies prioritize compatibility to existing code context over robustness, whereas tasks with broader file-level dependencies elicit additional reasoning steps, such as edge case handling and complete code generation, to ensure correctness. Overall, LRM reasoning is context-sensitive, adjusting the depth and type of actions according to task complexity and dependency structure.}

\subsubsection{Key Observations and Insights}
\label{sub:observation}
We further analyze the reasoning behaviors of LRMs and highlight both the positive and negative aspects of specific patterns.

\textbf{Preference in library selection.} Within the \KnowledgeRecall (Knowledge Recall) action, LRMs may propose solutions using libraries that are more efficient than naive approaches. For example:

\example{But the regex approach might be more efficient, especially if there are many keys. Because the loop approach would scan the entire string for each key, whereas the regex can find all placeholders in one pass. ...}{QWQ-32B on Task 62ece4982e6aefcf4aabbd62}

\noindent However, this behavior has limitations. LRMs may misidentify the required library due to ambiguities in the docstring or confusion between Python libraries/functions with overlapping names or functionality. Additionally, LRMs may lack up-to-date knowledge of certain libraries, or in some cases may avoid using libraries altogether when none are explicitly specified in the prompts.

\begin{figure}
    \includegraphics[width=0.99\columnwidth]{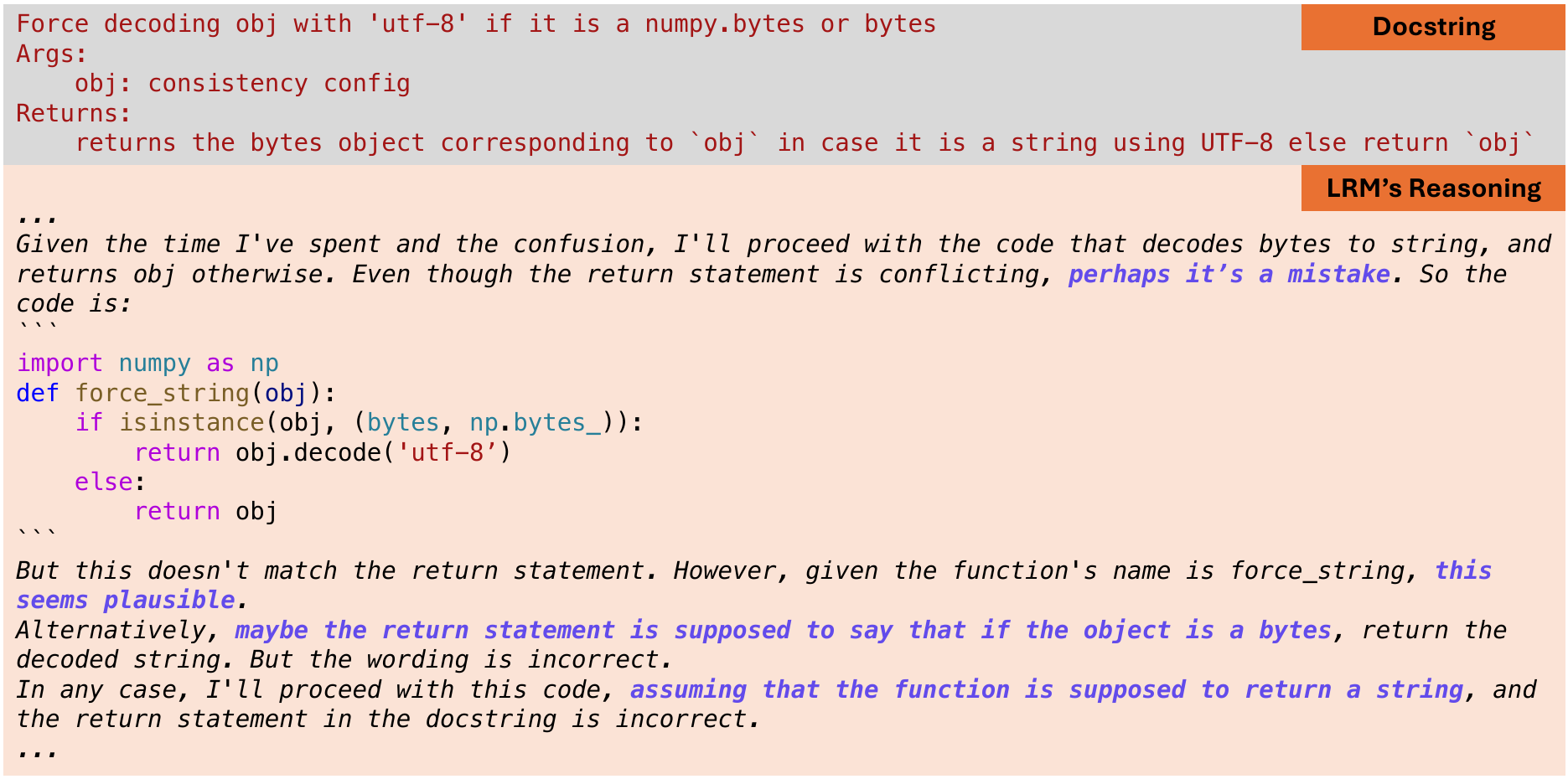}
    \vspace{-3mm}
    \caption{An example of harmful ambiguity recognition: Qwen3-1.7B becomes stuck in a loop of clarifying ambiguities and making assumptions (see \textcolor[rgb]{0.38,0.29,0.92}{\textbf{highlighted}} sentences).}
    \label{fig:LRM_AMB}
    \vspace{-3mm}
\end{figure}

\textbf{Ambiguity recognition and reliability of assumptions.} LRMs can detect missing or ambiguous information in prompts, as shown by their high frequency of Ambiguity Recognition (\AmbiguityRecognition) on \textit{class-runnable} (82.18\%) and \textit{file-runnable} (83.24\%) tasks. These tasks often lack necessary contextual details in the docstring. However, ambiguity can also harm reasoning quality. As shown in Figure \ref{fig:LRM_AMB}, unclear object types in input and return statements caused the LRM to fall into repetitive reasoning cycles without making progress.

\textbf{Consistency in the implementation phase.} The Implementation phase, which includes both scaffold and complete code generation, is the only phase not consistently present in reasoning traces, appearing in only 30\% to 55\% of cases across all dependency levels.. In some straightforward tasks, the LRMs skips scaffold or concrete code generation and move directly to cthe onclusion or review. Another observation is that the Complete Code Generation (\CompleteCodeGeneration) action does not always align with the final output. Reasoning traces may contain documentation-like explanations that are omitted in the final output, partly because the study's prompt explicitly required outputting only code. This highlights the need for caution when analyzing reasoning traces, since they may not fully correspond to the generated output.

\begin{figure}
    \includegraphics[width=0.99\columnwidth]{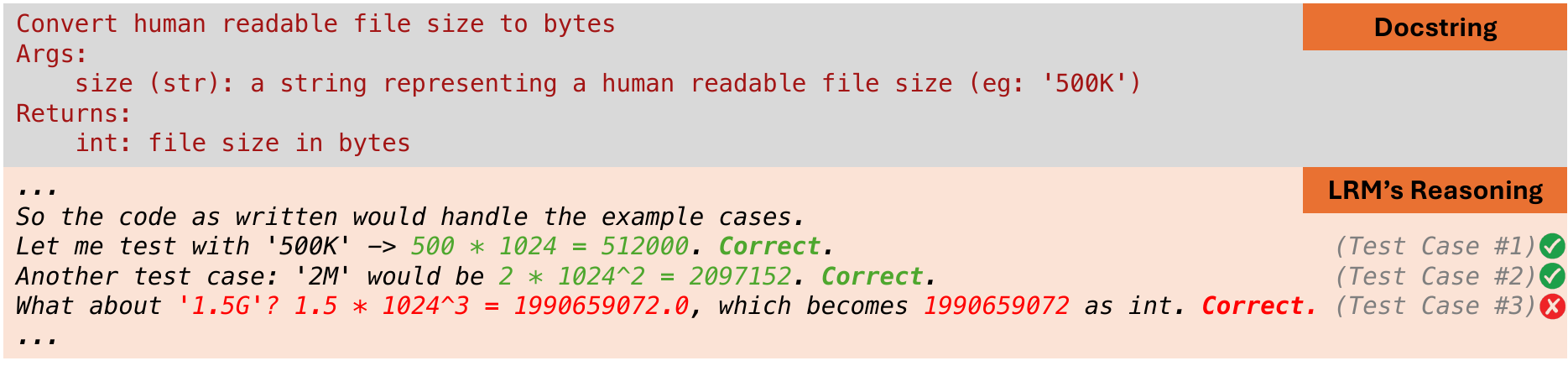}
    \vspace{-3mm}
    \caption{An example of unreliable unit test creation: Qwen3-14B generates three test cases and deems them all passed by its generated code, but test case \#3 contains an incorrect expected output.}
    \label{fig:LRM_UTC}
    \vspace{-3mm}
\end{figure}

\textbf{Test case creation and reliability issues.} LRMs display software-engineering-like behavior by generating unit tests for their own code without explicit prompting. However, this behavior is inconsistent. LRMs occasionally generate test cases for lower-dependency tasks, such as \textit{self-contained} (40\%) and \textit{slib-runnable} (32\%) levels, but they rarely do so for higher-dependency tasks, where ambiguity and missing context make test construction more difficult ($<11\%$). More severely, when test cases are created, their correctness is unreliable. In Figure \ref{fig:LRM_UTC}, the LRM generates test cases but fails to compute the correct expected output for test case \#3. While it constructs the correct reasoning steps, the final computation is flawed. This aligns with prior work identifying LRMs' struggles with mathematical reasoning under direct prompting \cite{cobbe2021training,gao2023pal}.

\begin{figure}
    \includegraphics[width=0.99\columnwidth]{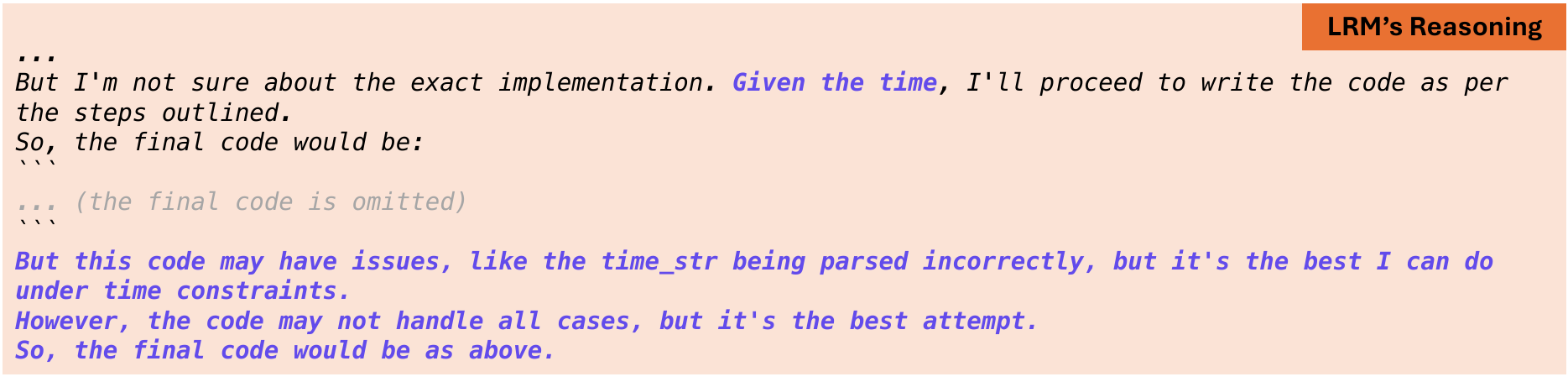}
    \vspace{-3mm}
    \caption{An example of invalid self-assertion: Qwen3-1.7B acknowledges potential issues but proceeds with the output nonetheless (see \textcolor[rgb]{0.38,0.29,0.92}{\textbf{highlighted}} sentences).}
    \label{fig:LRM_SFA}
    \vspace{-3mm}
\end{figure}

\textbf{Self-assertion and model confidence.} LRMs consistently exhibit self-assertion (\SelfAssertion), affirming that their generated code satisfies the task requirements. However, this does not always reflect genuine confidence. As illustrated in Figure \ref{fig:LRM_SFA}, LRMs sometimes recognize flaws during reasoning but, due to constraints such as limited context length or reasoning complexity, still output the flawed code. This mirrors human developers who, under time pressure, may knowingly submit imperfect solutions \cite{austin2001effects}.

\finding{
LRMs demonstrate a range of reasoning behaviors that mirror aspects of human problem solving, including efficient library use, ambiguity recognition, test case creation, and self-assertion. However, these behaviors are inconsistent: library choices may be outdated or incorrect, ambiguity often leads to flawed assumptions, implementation phases can be skipped or misaligned with outputs, and generated test cases are frequently unreliable. Self-assertion is consistently present but does not guarantee genuine confidence, as LRMs may still output flawed solutions. Together, these findings highlight both the adaptability and the current limitations of LRMs in performing reliable reasoning and producing correct code.
}

\subsection{RQ4: Prompting-based Improvements}
\label{sub:rq4}
We further evaluate the feasibility of two lightweight prompting-based strategies as potential improvements over our initial prompt, motivated by the key findings from the previous section:
\begin{itemize}[leftmargin=20pt]
    \item \textit{GUIDE}: Our previous analysis identifies Unit Test Creation (\UnitTestCreation) as the most positively correlated action with test passing. Based on this, we design a modified prompt that explicitly guides LRMs to emulate unit test cases, as shown in Figure~\ref{fig:guide}.
    \item \textit{CONT}: Since LRMs often struggle with ambiguity due to insufficient context, we test whether providing additional contextual information can improve performance. The modified prompt is illustrated in Figure~\ref{fig:cont}.
    
\end{itemize}

Table~\ref{tab:mutation} compares the Pass@1 scores of the modified prompts against the original prompt. Overall, both modified prompts yield slight improvements across most LRMs, with the exception of the Qwen3-14B variant on the \textit{CONT} prompting. For more complex tasks with higher dependencies, such as at the \textit{project-runnable} level, incorporating guidelines helps LRMs generate better code, resulting in improved performance. The consistent gains across different LRM series and parameter sizes indicate that integrating reasoning patterns into prompts holds promise as an effective strategy in prompt engineering and merits further investigation. However, due to the intrinsic randomness of LRMs, we cannot conclusively claim that these prompting strategies will always improve performance.

\begin{figure}
    \centering
    \footnotesize
    \begin{subfigure}[b]{0.5\textwidth} 
        \fbox{\begin{minipage}[t][0.15\textheight][t]{0.95\linewidth}
        You are a Python software engineer.
        \newline
        Generate Python code based on the following function signature and docstring. \newline
        \textbf{Do NOT include any explanation, reasoning, or markdown formatting.}\newline
        Output ONLY the code generated, in python markdown format.\newline
        
        \hlyellow{\#\# Tips} \newline
        \hlyellow{You should follow a test-driven development approach, first generating comprehensive unit tests before writing the actual code.} \newline
        
        [Function Signature] \newline
        [Human Docstring]
        \vfill
        \end{minipage}}
        \vspace{-1mm}
        \caption{\textit{\Placenametwo} prompt template.}\label{fig:cont}
    \end{subfigure}
    \hfill
    \begin{subfigure}[b]{0.48\textwidth}
        \fbox{\begin{minipage}[t][0.15\textheight][t]{0.95\linewidth}
        You are a Python software engineer.
        \newline
        Generate Python code based on the following function signature and docstring.\newline 
        Do NOT include any explanation, reasoning, or markdown formatting.\newline
        Output ONLY the code generated, in python markdown format.\newline
        
        \hlyellow{\#\# Context}\newline \hlyellow{Imported Packages: [Package dependencies]}\newline \hlyellow{Within file: [File dependencies]} \newline \hlyellow{Within class: [Class dependencies]} \newline

        [Function Signature] \newline
        [Human Docstring]
        \vfill
        \end{minipage}}
        \vspace{-1mm}
        \caption{\textit{\Placenameone} prompt template.}\label{fig:guide}
    \end{subfigure}
    \vspace{-4mm}
    \caption{Prompt templates for the two lightweight prompting methods, with newly added lines highlighted and lines removed are bolded.}
    \vspace{-6mm}
\end{figure}

\begin{table*}
\centering
\caption{Pass@1 score for code generation task (\%) with and without prompt modifications.}
\label{tab:mutation}
\vspace{-3mm}
\resizebox{\linewidth}{!}{%
\begin{tabular}{l ccc ccc ccc ccc ccc}
\toprule
\textbf{Category} 
& \multicolumn{3}{c}{\textbf{DeepSeek-R1-7B}} 
& \multicolumn{3}{c}{\textbf{Qwen3-1.7B}} 
& \multicolumn{3}{c}{\textbf{Qwen3-8B}} 
& \multicolumn{3}{c}{\textbf{Qwen3-14B}} 
& \multicolumn{3}{c}{\textbf{QWQ-32B 32B}} \\
\cmidrule(lr){2-4} \cmidrule(lr){5-7} \cmidrule(lr){8-10} \cmidrule(lr){11-13} \cmidrule(lr){14-16}
& Original & \Placenameone & \textit{\Placenametwo} 
& Original & \Placenameone & \textit{\Placenametwo} 
& Original & \Placenameone & \textit{\Placenametwo} 
& Original & \Placenameone & \textit{\Placenametwo} 
& Original & \Placenameone & \textit{\Placenametwo}   \\
\midrule
\textbf{\textit{self-contained}}     & 28.57 & 37.14 & 28.57 & 40.00 & 40.00 & 34.29 & 51.43 & 42.86 & 54.29 & 51.43 & 48.57 & 57.14 & 54.29 & 51.43 & 54.29 \\
\textbf{\textit{slib-runnable}}      & 25.00 & 25.00 & 21.43 & 39.29 & 28.57 & 32.14 & 50.00 & 50.00 & 46.43 & 50.00 & 39.29 & 50.00 & 46.43 & 53.57 & 50.00 \\
\textbf{\textit{plib-runnable}}      & 19.05 & 19.05 & 19.05 & 19.05 & 23.81 & 19.05 & 28.57 & 23.81 & 33.33 & 28.57 & 38.10 & 33.33 & 38.1 & 33.33 & 38.10 \\
\textbf{\textit{class-runnable}}     & 10.91 & 12.73 & 14.55 & 12.73 & 23.64 & 20.00 & 23.64 & 23.64 & 21.82 & 23.64 & 23.64 & 25.45 & 18.52 & 20.00 & 16.36 \\
\textbf{\textit{file-runnable}}      & 10.29 & 10.29 & 10.29 & 17.65 & 16.18 & 19.12 & 25.00 & 30.88 & 22.06 & 25.00 & 26.47 & 25.00 & 19.12 & 23.53 & 22.06 \\
\textbf{\textit{project-runnable}}   & 4.35 & 0.00 & 4.35 & 4.35 & 8.70 & 8.70 & 4.35 & 8.70 &  13.04 & 13.04 & 8.70 & 13.04 & 4.35 & 13.04 & 17.39 \\

\midrule
\textbf{overall}            & 15.22 & \hlgreen{16.52} & \hlgreen{15.65} & 21.30 & \hlgreen{23.04} & \hlgreen{22.17} & 29.57 & \hlgreen{30.43} & \hlgreen{30.00} & 30.87 & \hlred{30.00} & \hlgreen{32.61} & 27.83 & \hlgreen{30.43} & \hlgreen{30.00} \\

\bottomrule
\end{tabular}%
}

\end{table*}

\finding{Our results highlight the potential of incorporating context or reasoning guidelines into prompts to enhance LRM-generated code. The \textit{GUIDE} prompt variant, which integrates a UTC-enhanced design, shows slight improvements across different LRMs.}
\section{Discussion}

\subsection{Implications for Researchers} 




\textbf{Reasoning Behavior Visualization.} Our study introduces a taxonomy of fine-grained reasoning actions, showing that LRMs exhibit distinct reasoning patterns across different tasks. To deepen understanding of these reasoning traces, future research can develop visualization techniques grounded in our taxonomy and annotated data, thereby helping to open the black box of model reasoning in code generation. Such visualizations can enhance the interpretability of LRMs and foster greater developer trust in their outputs.

\noindent\textbf{Reasoning Capability Enhancement.} Our study investigates the feasibility of prompting-based improvement strategies, showing that while these methods yield modest gains, there remains considerable room for advancement. Future research can explore more sophisticated approaches to enhancing reasoning capabilities. For instance, we observe that DeepSeek-R1 tends to adopt a linear, waterfall-style reasoning process, whereas Qwen3 models favor an iterative reasoning process, achieving higher functional correctness. This suggests the potential of generating higher-quality reasoning traces that emulate realistic iterative programming practices for model fine-tuning. Likewise, the strong positive correlation between UTC and test pass rates highlights the promise of incorporating the test-driven development (TDD) paradigm into the LRMs for code generation.

\subsection{Implication for Software Developers} 
\textbf{Prompt/Context Engineering for LRMs.} Since LRMs inherently exhibit CoT-like reasoning, some argue that carefully crafted prompts are unnecessary when using them. However, our study suggests that prompt engineering remains important for effective code generation. When designing prompts, practitioners should provide concise instructions while ensuring that the problem statement and contextual information, such as well-written docstrings, relevant libraries, classes, or attributes, are explicit and complete. Our findings show that incomplete, unclear, or ambiguous prompts can hinder LRMs' reasoning and negatively impact performance. This insight also highlights the emerging role of context engineering, which focuses on supplying precise and comprehensive contextual information tailored to the task requirements.

\noindent\textbf{Reasoning Process Inspection}. Our study provides insights into the limitations of LRMs’ reasoning in code generation. For instance, LRMs still struggle to reliably perform tasks such as generating tests or maintaining sound assumptions. This indicates that developers should not only carefully review and validate the code produced by LRMs for potential flaws but also examine and verify the models’ reasoning traces.

\section{Threats to Validity}

\textbf{External Validity.}  
Threats to external validity concern the extent to which our findings generalize across different languages. Due to the imbalanced distribution of dependency scopes in Java, we focused our analysis on reasoning traces in Python. Exploring reasoning traces for other programming languages remains an important direction for future work. Another potential threat is the quality of code generation tasks considered. We mitigated this by using CoderEval, a widely adopted dataset of real-world programming tasks that spans multiple dependency levels.  

\noindent\textbf{Internal Validity.}  
Threats to internal validity mainly relate to manual annotation and taxonomy construction. To address this, we applied standard conflict-resolution strategies during taxonomy development to ensure reliability. Additionally, since the model temperature is non-zero, as recommended by the LRM publishers \cite{deepseek_hf}, outputs are inherently nondeterministic. In our experiments, setting the temperature to zero led to inconsistent behavior, including timeouts and incoherent repetitive reasoning. To mitigate potential reproducibility concerns, we documented all model responses and outputs in our replication package.


\section{Related Work}

\textbf{LLMs in Code Generation.}
Large language models (LLMs) have been increasingly applied to software engineering tasks, particularly code generation~\cite{fakhoury2024llm,zheng2024opencodeinterpreter, zhang2024codeagentenhancingcodegeneration,li2023acecoder,tong-zhang-2024-codejudge,gu2023llm}.
Existing literature~\cite{liu2023your, chatziveroglou2025exploring, mu2024clarifygpt, huang2024knowledge} has primarily focused on the GPT series, most notably GPT-4 as the primary benchmark for performance evaluation~\cite{hou2024largelanguagemodelssoftware}. Alongside general-purpose models, several LLMs have been explicitly trained for code-related tasks, such as Code Llama ~\cite{roziere2023code} and Qwen2.5-Coder~\cite{hui2024qwen25codertechnicalreport}. To assess these models, various datasets have been created, such as HumanEval~\cite{chen2021evaluating}, MBPP~\cite{austin2021program}, ClassEval~\cite{du2023classeval}, DevEval~\cite{li2024deveval}, and CoderEval~\cite{yu2024codereval}.


\noindent\textbf{LLM Reasoning.}
With the emergence of large reasoning models (LRMs) and greater transparency enabled by their explicit intermediate reasoning steps, there has been a growing interest in analysing the models’ reasoning traces to understand their internal decision-making processes better.
Recent studies have sought to categorise the reasoning behaviour of large language models across various domains. Marjanovic et al. \cite{marjanovic2025deepseek}, inspired by common human reasoning, derive a general taxonomy of DeepSeek-R1 reasoning patterns on a diverse set of problems. In their study, Ming et al. \cite{ming2025helpful} explore and derive a taxonomy on LLM reasoning behavior as a critique and when faced with critique. Plaat et al. \cite{plaat2024reasoning} conduct a survey and propose a 3-stage taxonomy on LLM reasoning for math problems. Bandyopadhyay et al. \cite{bandyopadhyay2025thinking} derives common phases on how LLM would reason as a pipeline/structure.
Several studies have directly addressed reasoning within the context of code generation. Wei et al.\cite{wei2025evaluating} investigate the reasoning model's capability in performing code generation on the competitive programming domain and highlight some flaws in the code generated by LLM. Liu et al. \cite{liu2024codemind} in their proposed framework evaluation CodeMind, highlight reasoning model capability and limitation on control flow, and show that code generation ability, despite being correlated, does not imply code reasoning.
Prompt variation has also been explored as a tool for influencing reasoning behavior. In their study, Chatziveroglou et al. \cite{chatziveroglou2025exploring} explored the effect on LLM reasoning through prompt variation on math problems. Mu et al. \cite{mu2024clarifygpt} in ClarifyGPT propose a framework to enhance LLM on code generation through ambiguity detection and clarification.
\section{Conclusion}
In this paper, we conduct an empirical study of reasoning behaviors in large reasoning models (LRMs) for code generation. We develop a taxonomy of 15 reasoning actions across four phases, reveal common and model-specific reasoning patterns, and show how these behaviors affect functional correctness. Finally, we demonstrate that lightweight, reasoning-oriented prompting strategies can further improve code generation. These findings advance our understanding of LRM reasoning and provide practical guidance for automated software development.

\section{Data Availability}
All of the artefacts and LLM output are made publicly available and can be accessed here:\\ \url{https://github.com/ReasoningPattern/ReasoningPattern}

\bibliographystyle{ACM-Reference-Format}
\bibliography{references}


\begin{thebibliography}{43}


\ifx \showCODEN    \undefined \def \showCODEN     #1{\unskip}     \fi
\ifx \showISBNx    \undefined \def \showISBNx     #1{\unskip}     \fi
\ifx \showISBNxiii \undefined \def \showISBNxiii  #1{\unskip}     \fi
\ifx \showISSN     \undefined \def \showISSN      #1{\unskip}     \fi
\ifx \showLCCN     \undefined \def \showLCCN      #1{\unskip}     \fi
\ifx \shownote     \undefined \def \shownote      #1{#1}          \fi
\ifx \showarticletitle \undefined \def \showarticletitle #1{#1}   \fi
\ifx \showURL      \undefined \def \showURL       {\relax}        \fi
\providecommand\bibfield[2]{#2}
\providecommand\bibinfo[2]{#2}
\providecommand\natexlab[1]{#1}
\providecommand\showeprint[2][]{arXiv:#2}

\bibitem[{Anthropic}(2025)]%
        {claude}
\bibfield{author}{\bibinfo{person}{{Anthropic}}.} \bibinfo{year}{2025}\natexlab{}.
\newblock \bibinfo{title}{Introducing Claude 4}.
\newblock
\urldef\tempurl%
\url{https://www.anthropic.com/news/claude-4}
\showURL{%
\tempurl}
\newblock
\shownote{Accessed July 2025}.


\bibitem[Austin et~al\mbox{.}(2021)]%
        {austin2021program}
\bibfield{author}{\bibinfo{person}{Jacob Austin}, \bibinfo{person}{Augustus Odena}, \bibinfo{person}{Maxwell Nye}, \bibinfo{person}{Maarten Bosma}, \bibinfo{person}{Henryk Michalewski}, \bibinfo{person}{David Dohan}, \bibinfo{person}{Ellen Jiang}, \bibinfo{person}{Carrie Cai}, \bibinfo{person}{Michael Terry}, \bibinfo{person}{Quoc Le}, {et~al\mbox{.}}} \bibinfo{year}{2021}\natexlab{}.
\newblock \showarticletitle{Program synthesis with large language models}.
\newblock \bibinfo{journal}{\emph{arXiv preprint arXiv:2108.07732}} (\bibinfo{year}{2021}).
\newblock


\bibitem[Austin(2001)]%
        {austin2001effects}
\bibfield{author}{\bibinfo{person}{Robert~D Austin}.} \bibinfo{year}{2001}\natexlab{}.
\newblock \showarticletitle{The effects of time pressure on quality in software development: An agency model}.
\newblock \bibinfo{journal}{\emph{Information systems research}} \bibinfo{volume}{12}, \bibinfo{number}{2} (\bibinfo{year}{2001}), \bibinfo{pages}{195--207}.
\newblock


\bibitem[Bandyopadhyay et~al\mbox{.}(2025)]%
        {bandyopadhyay2025thinking}
\bibfield{author}{\bibinfo{person}{Dibyanayan Bandyopadhyay}, \bibinfo{person}{Soham Bhattacharjee}, {and} \bibinfo{person}{Asif Ekbal}.} \bibinfo{year}{2025}\natexlab{}.
\newblock \showarticletitle{Thinking machines: A survey of llm based reasoning strategies}.
\newblock \bibinfo{journal}{\emph{arXiv preprint arXiv:2503.10814}} (\bibinfo{year}{2025}).
\newblock


\bibitem[Besta et~al\mbox{.}(2025)]%
        {bestaReasoningLanguageModels2025}
\bibfield{author}{\bibinfo{person}{Maciej Besta}, \bibinfo{person}{Julia Barth}, \bibinfo{person}{Eric Schreiber}, \bibinfo{person}{Ales Kubicek}, \bibinfo{person}{Afonso Catarino}, \bibinfo{person}{Robert Gerstenberger}, \bibinfo{person}{Piotr Nyczyk}, \bibinfo{person}{Patrick Iff}, \bibinfo{person}{Yueling Li}, \bibinfo{person}{Sam Houliston}, \bibinfo{person}{Tomasz Sternal}, \bibinfo{person}{Marcin Copik}, \bibinfo{person}{Grzegorz Kwaśniewski}, \bibinfo{person}{Jürgen Müller}, \bibinfo{person}{Łukasz Flis}, \bibinfo{person}{Hannes Eberhard}, \bibinfo{person}{Zixuan Chen}, \bibinfo{person}{Hubert Niewiadomski}, {and} \bibinfo{person}{Torsten Hoefler}.} \bibinfo{year}{2025}\natexlab{}.
\newblock \bibinfo{title}{Reasoning {Language} {Models}: {A} {Blueprint}}.
\newblock
\href{https://doi.org/10.48550/arXiv.2501.11223}{doi:\nolinkurl{10.48550/arXiv.2501.11223}}
\newblock
\shownote{arXiv:2501.11223 [cs]}.


\bibitem[Chatziveroglou et~al\mbox{.}(2025)]%
        {chatziveroglou2025exploring}
\bibfield{author}{\bibinfo{person}{Giannis Chatziveroglou}, \bibinfo{person}{Richard Yun}, {and} \bibinfo{person}{Maura Kelleher}.} \bibinfo{year}{2025}\natexlab{}.
\newblock \showarticletitle{Exploring llm reasoning through controlled prompt variations}.
\newblock \bibinfo{journal}{\emph{arXiv preprint arXiv:2504.02111}} (\bibinfo{year}{2025}).
\newblock


\bibitem[Chen et~al\mbox{.}(2021)]%
        {chen2021evaluating}
\bibfield{author}{\bibinfo{person}{Mark Chen}, \bibinfo{person}{Jerry Tworek}, \bibinfo{person}{Heewoo Jun}, \bibinfo{person}{Qiming Yuan}, \bibinfo{person}{Henrique Ponde De~Oliveira Pinto}, \bibinfo{person}{Jared Kaplan}, \bibinfo{person}{Harri Edwards}, \bibinfo{person}{Yuri Burda}, \bibinfo{person}{Nicholas Joseph}, \bibinfo{person}{Greg Brockman}, {et~al\mbox{.}}} \bibinfo{year}{2021}\natexlab{}.
\newblock \showarticletitle{Evaluating large language models trained on code}.
\newblock \bibinfo{journal}{\emph{arXiv preprint arXiv:2107.03374}} (\bibinfo{year}{2021}).
\newblock


\bibitem[Cobbe et~al\mbox{.}(2021)]%
        {cobbe2021training}
\bibfield{author}{\bibinfo{person}{Karl Cobbe}, \bibinfo{person}{Vineet Kosaraju}, \bibinfo{person}{Mohammad Bavarian}, \bibinfo{person}{Mark Chen}, \bibinfo{person}{Heewoo Jun}, \bibinfo{person}{Lukasz Kaiser}, \bibinfo{person}{Matthias Plappert}, \bibinfo{person}{Jerry Tworek}, \bibinfo{person}{Jacob Hilton}, \bibinfo{person}{Reiichiro Nakano}, {et~al\mbox{.}}} \bibinfo{year}{2021}\natexlab{}.
\newblock \showarticletitle{Training verifiers to solve math word problems}.
\newblock \bibinfo{journal}{\emph{arXiv preprint arXiv:2110.14168}} (\bibinfo{year}{2021}).
\newblock


\bibitem[Cohen(1960)]%
        {cohen1960coefficient}
\bibfield{author}{\bibinfo{person}{Jacob Cohen}.} \bibinfo{year}{1960}\natexlab{}.
\newblock \showarticletitle{A coefficient of agreement for nominal scales}.
\newblock \bibinfo{journal}{\emph{Educational and psychological measurement}} \bibinfo{volume}{20}, \bibinfo{number}{1} (\bibinfo{year}{1960}), \bibinfo{pages}{37--46}.
\newblock


\bibitem[Cram{\'e}r(1999)]%
        {cramer1999mathematical}
\bibfield{author}{\bibinfo{person}{Harald Cram{\'e}r}.} \bibinfo{year}{1999}\natexlab{}.
\newblock \bibinfo{booktitle}{\emph{Mathematical methods of statistics}}. Vol.~\bibinfo{volume}{9}.
\newblock \bibinfo{publisher}{Princeton university press}.
\newblock


\bibitem[DeepSeek-AI(2025)]%
        {deepseekai2025deepseekr1incentivizingreasoningcapability}
\bibfield{author}{\bibinfo{person}{DeepSeek-AI}.} \bibinfo{year}{2025}\natexlab{}.
\newblock \bibinfo{title}{DeepSeek-R1: Incentivizing Reasoning Capability in LLMs via Reinforcement Learning}.
\newblock
\showeprint[arxiv]{2501.12948}~[cs.CL]
\urldef\tempurl%
\url{https://arxiv.org/abs/2501.12948}
\showURL{%
\tempurl}


\bibitem[{DeepSeek-AI}(2025)]%
        {deepseek_hf}
\bibfield{author}{\bibinfo{person}{{DeepSeek-AI}}.} \bibinfo{year}{2025}\natexlab{}.
\newblock \bibinfo{title}{DeepSeek-R1 Model on Hugging Face}.
\newblock
\urldef\tempurl%
\url{https://huggingface.co/deepseek-ai/DeepSeek-R1}
\showURL{%
\tempurl}
\newblock
\shownote{Accessed July 2025}.


\bibitem[Du et~al\mbox{.}(2023)]%
        {du2023classeval}
\bibfield{author}{\bibinfo{person}{Xueying Du}, \bibinfo{person}{Mingwei Liu}, \bibinfo{person}{Kaixin Wang}, \bibinfo{person}{Hanlin Wang}, \bibinfo{person}{Junwei Liu}, \bibinfo{person}{Yixuan Chen}, \bibinfo{person}{Jiayi Feng}, \bibinfo{person}{Chaofeng Sha}, \bibinfo{person}{Xin Peng}, {and} \bibinfo{person}{Yiling Lou}.} \bibinfo{year}{2023}\natexlab{}.
\newblock \showarticletitle{Classeval: A manually-crafted benchmark for evaluating llms on class-level code generation}.
\newblock \bibinfo{journal}{\emph{arXiv preprint arXiv:2308.01861}} (\bibinfo{year}{2023}).
\newblock


\bibitem[Fakhoury et~al\mbox{.}(2024)]%
        {fakhoury2024llm}
\bibfield{author}{\bibinfo{person}{Sarah Fakhoury}, \bibinfo{person}{Aaditya Naik}, \bibinfo{person}{Georgios Sakkas}, \bibinfo{person}{Saikat Chakraborty}, {and} \bibinfo{person}{Shuvendu~K Lahiri}.} \bibinfo{year}{2024}\natexlab{}.
\newblock \showarticletitle{Llm-based test-driven interactive code generation: User study and empirical evaluation}.
\newblock \bibinfo{journal}{\emph{IEEE Transactions on Software Engineering}} (\bibinfo{year}{2024}).
\newblock


\bibitem[Friedman(2022)]%
        {Friedman_2022}
\bibfield{author}{\bibinfo{person}{Nat Friedman}.} \bibinfo{year}{2022}\natexlab{}.
\newblock \bibinfo{title}{Introducing github copilot: Your AI pair programmer}.
\newblock
\urldef\tempurl%
\url{https://github.blog/news-insights/product-news/introducing-github-copilot-ai-pair-programmer/}
\showURL{%
\tempurl}


\bibitem[Gao et~al\mbox{.}(2023)]%
        {gao2023pal}
\bibfield{author}{\bibinfo{person}{Luyu Gao}, \bibinfo{person}{Aman Madaan}, \bibinfo{person}{Shuyan Zhou}, \bibinfo{person}{Uri Alon}, \bibinfo{person}{Pengfei Liu}, \bibinfo{person}{Yiming Yang}, \bibinfo{person}{Jamie Callan}, {and} \bibinfo{person}{Graham Neubig}.} \bibinfo{year}{2023}\natexlab{}.
\newblock \showarticletitle{Pal: Program-aided language models}. In \bibinfo{booktitle}{\emph{International Conference on Machine Learning}}. PMLR, \bibinfo{pages}{10764--10799}.
\newblock


\bibitem[Gu(2023)]%
        {gu2023llm}
\bibfield{author}{\bibinfo{person}{Qiuhan Gu}.} \bibinfo{year}{2023}\natexlab{}.
\newblock \showarticletitle{Llm-based code generation method for golang compiler testing}. In \bibinfo{booktitle}{\emph{Proceedings of the 31st ACM Joint European Software Engineering Conference and Symposium on the Foundations of Software Engineering}}. \bibinfo{pages}{2201--2203}.
\newblock


\bibitem[Hou et~al\mbox{.}(2024)]%
        {hou2024largelanguagemodelssoftware}
\bibfield{author}{\bibinfo{person}{Xinyi Hou}, \bibinfo{person}{Yanjie Zhao}, \bibinfo{person}{Yue Liu}, \bibinfo{person}{Zhou Yang}, \bibinfo{person}{Kailong Wang}, \bibinfo{person}{Li Li}, \bibinfo{person}{Xiapu Luo}, \bibinfo{person}{David Lo}, \bibinfo{person}{John Grundy}, {and} \bibinfo{person}{Haoyu Wang}.} \bibinfo{year}{2024}\natexlab{}.
\newblock \bibinfo{title}{Large Language Models for Software Engineering: A Systematic Literature Review}.
\newblock
\showeprint[arxiv]{2308.10620}~[cs.SE]
\urldef\tempurl%
\url{https://arxiv.org/abs/2308.10620}
\showURL{%
\tempurl}


\bibitem[Huang et~al\mbox{.}(2000)]%
        {huang2000fast}
\bibfield{author}{\bibinfo{person}{Liusheng Huang}, \bibinfo{person}{Huaping Chen}, \bibinfo{person}{Xun Wang}, {and} \bibinfo{person}{Guoliang Chen}.} \bibinfo{year}{2000}\natexlab{}.
\newblock \showarticletitle{A fast algorithm for mining association rules}.
\newblock \bibinfo{journal}{\emph{Journal of Computer Science and Technology}} \bibinfo{volume}{15}, \bibinfo{number}{6} (\bibinfo{year}{2000}), \bibinfo{pages}{619--624}.
\newblock


\bibitem[Huang et~al\mbox{.}(2024)]%
        {huang2024knowledge}
\bibfield{author}{\bibinfo{person}{Tao Huang}, \bibinfo{person}{Zhihong Sun}, \bibinfo{person}{Zhi Jin}, \bibinfo{person}{Ge Li}, {and} \bibinfo{person}{Chen Lyu}.} \bibinfo{year}{2024}\natexlab{}.
\newblock \showarticletitle{Knowledge-aware code generation with large language models}. In \bibinfo{booktitle}{\emph{Proceedings of the 32nd IEEE/ACM International Conference on Program Comprehension}}. \bibinfo{pages}{52--63}.
\newblock


\bibitem[Hui et~al\mbox{.}(2024)]%
        {hui2024qwen25codertechnicalreport}
\bibfield{author}{\bibinfo{person}{Binyuan Hui}, \bibinfo{person}{Jian Yang}, \bibinfo{person}{Zeyu Cui}, \bibinfo{person}{Jiaxi Yang}, \bibinfo{person}{Dayiheng Liu}, \bibinfo{person}{Lei Zhang}, \bibinfo{person}{Tianyu Liu}, \bibinfo{person}{Jiajun Zhang}, \bibinfo{person}{Bowen Yu}, \bibinfo{person}{Keming Lu}, \bibinfo{person}{Kai Dang}, \bibinfo{person}{Yang Fan}, \bibinfo{person}{Yichang Zhang}, \bibinfo{person}{An Yang}, \bibinfo{person}{Rui Men}, \bibinfo{person}{Fei Huang}, \bibinfo{person}{Bo Zheng}, \bibinfo{person}{Yibo Miao}, \bibinfo{person}{Shanghaoran Quan}, \bibinfo{person}{Yunlong Feng}, \bibinfo{person}{Xingzhang Ren}, \bibinfo{person}{Xuancheng Ren}, \bibinfo{person}{Jingren Zhou}, {and} \bibinfo{person}{Junyang Lin}.} \bibinfo{year}{2024}\natexlab{}.
\newblock \bibinfo{title}{Qwen2.5-Coder Technical Report}.
\newblock
\showeprint[arxiv]{2409.12186}~[cs.CL]
\urldef\tempurl%
\url{https://arxiv.org/abs/2409.12186}
\showURL{%
\tempurl}


\bibitem[Ji et~al\mbox{.}(2023)]%
        {ji2023survey}
\bibfield{author}{\bibinfo{person}{Ziwei Ji}, \bibinfo{person}{Nayeon Lee}, \bibinfo{person}{Rita Frieske}, \bibinfo{person}{Tiezheng Yu}, \bibinfo{person}{Dan Su}, \bibinfo{person}{Yan Xu}, \bibinfo{person}{Etsuko Ishii}, \bibinfo{person}{Ye~Jin Bang}, \bibinfo{person}{Andrea Madotto}, {and} \bibinfo{person}{Pascale Fung}.} \bibinfo{year}{2023}\natexlab{}.
\newblock \showarticletitle{Survey of hallucination in natural language generation}.
\newblock \bibinfo{journal}{\emph{ACM computing surveys}} \bibinfo{volume}{55}, \bibinfo{number}{12} (\bibinfo{year}{2023}), \bibinfo{pages}{1--38}.
\newblock


\bibitem[Khandkar(2009)]%
        {opencoding}
\bibfield{author}{\bibinfo{person}{Shahedul~Huq Khandkar}.} \bibinfo{year}{2009}\natexlab{}.
\newblock \showarticletitle{Open coding}.
\newblock \bibinfo{journal}{\emph{University of Calgary}} \bibinfo{volume}{23}, \bibinfo{number}{2009} (\bibinfo{year}{2009}), \bibinfo{pages}{2009}.
\newblock


\bibitem[Landis and Koch(1977)]%
        {landis1977measurement}
\bibfield{author}{\bibinfo{person}{J~Richard Landis} {and} \bibinfo{person}{Gary~G Koch}.} \bibinfo{year}{1977}\natexlab{}.
\newblock \showarticletitle{The measurement of observer agreement for categorical data}.
\newblock \bibinfo{journal}{\emph{biometrics}} (\bibinfo{year}{1977}), \bibinfo{pages}{159--174}.
\newblock


\bibitem[Li et~al\mbox{.}(2024)]%
        {li2024deveval}
\bibfield{author}{\bibinfo{person}{Jia Li}, \bibinfo{person}{Ge Li}, \bibinfo{person}{Yunfei Zhao}, \bibinfo{person}{Yongmin Li}, \bibinfo{person}{Huanyu Liu}, \bibinfo{person}{Hao Zhu}, \bibinfo{person}{Lecheng Wang}, \bibinfo{person}{Kaibo Liu}, \bibinfo{person}{Zheng Fang}, \bibinfo{person}{Lanshen Wang}, {et~al\mbox{.}}} \bibinfo{year}{2024}\natexlab{}.
\newblock \showarticletitle{Deveval: A manually-annotated code generation benchmark aligned with real-world code repositories}.
\newblock \bibinfo{journal}{\emph{arXiv preprint arXiv:2405.19856}} (\bibinfo{year}{2024}).
\newblock


\bibitem[Li et~al\mbox{.}(2023)]%
        {li2023acecoder}
\bibfield{author}{\bibinfo{person}{Jia Li}, \bibinfo{person}{Yunfei Zhao}, \bibinfo{person}{Yongmin Li}, \bibinfo{person}{Ge Li}, {and} \bibinfo{person}{Zhi Jin}.} \bibinfo{year}{2023}\natexlab{}.
\newblock \showarticletitle{Acecoder: Utilizing existing code to enhance code generation}.
\newblock \bibinfo{journal}{\emph{arXiv preprint arXiv:2303.17780}} (\bibinfo{year}{2023}).
\newblock


\bibitem[Liu et~al\mbox{.}(2024)]%
        {liu2024codemind}
\bibfield{author}{\bibinfo{person}{Changshu Liu}, \bibinfo{person}{Yang Chen}, {and} \bibinfo{person}{Reyhaneh Jabbarvand}.} \bibinfo{year}{2024}\natexlab{}.
\newblock \showarticletitle{CodeMind: Evaluating Large Language Models for Code Reasoning}.
\newblock \bibinfo{journal}{\emph{arXiv preprint arXiv:2402.09664}} (\bibinfo{year}{2024}).
\newblock


\bibitem[Liu et~al\mbox{.}(2023)]%
        {liu2023your}
\bibfield{author}{\bibinfo{person}{Jiawei Liu}, \bibinfo{person}{Chunqiu~Steven Xia}, \bibinfo{person}{Yuyao Wang}, {and} \bibinfo{person}{Lingming Zhang}.} \bibinfo{year}{2023}\natexlab{}.
\newblock \showarticletitle{Is your code generated by chatgpt really correct? rigorous evaluation of large language models for code generation}.
\newblock \bibinfo{journal}{\emph{Advances in Neural Information Processing Systems}}  \bibinfo{volume}{36} (\bibinfo{year}{2023}), \bibinfo{pages}{21558--21572}.
\newblock


\bibitem[Marjanovi{\'c} et~al\mbox{.}(2025)]%
        {marjanovic2025deepseek}
\bibfield{author}{\bibinfo{person}{Sara~Vera Marjanovi{\'c}}, \bibinfo{person}{Arkil Patel}, \bibinfo{person}{Vaibhav Adlakha}, \bibinfo{person}{Milad Aghajohari}, \bibinfo{person}{Parishad BehnamGhader}, \bibinfo{person}{Mehar Bhatia}, \bibinfo{person}{Aditi Khandelwal}, \bibinfo{person}{Austin Kraft}, \bibinfo{person}{Benno Krojer}, \bibinfo{person}{Xing~Han L{\`u}}, {et~al\mbox{.}}} \bibinfo{year}{2025}\natexlab{}.
\newblock \showarticletitle{DeepSeek-R1 Thoughtology: Let's think about LLM Reasoning}.
\newblock \bibinfo{journal}{\emph{arXiv preprint arXiv:2504.07128}} (\bibinfo{year}{2025}).
\newblock


\bibitem[Ming et~al\mbox{.}(2025)]%
        {ming2025helpful}
\bibfield{author}{\bibinfo{person}{Yifei Ming}, \bibinfo{person}{Zixuan Ke}, \bibinfo{person}{Xuan-Phi Nguyen}, \bibinfo{person}{Jiayu Wang}, {and} \bibinfo{person}{Shafiq Joty}.} \bibinfo{year}{2025}\natexlab{}.
\newblock \showarticletitle{Helpful Agent Meets Deceptive Judge: Understanding Vulnerabilities in Agentic Workflows}.
\newblock \bibinfo{journal}{\emph{arXiv preprint arXiv:2506.03332}} (\bibinfo{year}{2025}).
\newblock


\bibitem[Mu et~al\mbox{.}(2024)]%
        {mu2024clarifygpt}
\bibfield{author}{\bibinfo{person}{Fangwen Mu}, \bibinfo{person}{Lin Shi}, \bibinfo{person}{Song Wang}, \bibinfo{person}{Zhuohao Yu}, \bibinfo{person}{Binquan Zhang}, \bibinfo{person}{ChenXue Wang}, \bibinfo{person}{Shichao Liu}, {and} \bibinfo{person}{Qing Wang}.} \bibinfo{year}{2024}\natexlab{}.
\newblock \showarticletitle{Clarifygpt: A framework for enhancing llm-based code generation via requirements clarification}.
\newblock \bibinfo{journal}{\emph{Proceedings of the ACM on Software Engineering}} \bibinfo{volume}{1}, \bibinfo{number}{FSE} (\bibinfo{year}{2024}), \bibinfo{pages}{2332--2354}.
\newblock


\bibitem[{OpenAI}(2025)]%
        {openai}
\bibfield{author}{\bibinfo{person}{{OpenAI}}.} \bibinfo{year}{2025}\natexlab{}.
\newblock \bibinfo{title}{Introducing OpenAI o3 and o4-mini}.
\newblock
\urldef\tempurl%
\url{https://openai.com/index/introducing-o3-and-o4-mini/}
\showURL{%
\tempurl}
\newblock
\shownote{Accessed July 2025}.


\bibitem[Plaat et~al\mbox{.}(2024)]%
        {plaat2024reasoning}
\bibfield{author}{\bibinfo{person}{Aske Plaat}, \bibinfo{person}{Annie Wong}, \bibinfo{person}{Suzan Verberne}, \bibinfo{person}{Joost Broekens}, \bibinfo{person}{Niki van Stein}, {and} \bibinfo{person}{Thomas Back}.} \bibinfo{year}{2024}\natexlab{}.
\newblock \showarticletitle{Reasoning with large language models, a survey}.
\newblock \bibinfo{journal}{\emph{arXiv preprint arXiv:2407.11511}} (\bibinfo{year}{2024}).
\newblock


\bibitem[Roziere et~al\mbox{.}(2023)]%
        {roziere2023code}
\bibfield{author}{\bibinfo{person}{Baptiste Roziere}, \bibinfo{person}{Jonas Gehring}, \bibinfo{person}{Fabian Gloeckle}, \bibinfo{person}{Sten Sootla}, \bibinfo{person}{Itai Gat}, \bibinfo{person}{Xiaoqing~Ellen Tan}, \bibinfo{person}{Yossi Adi}, \bibinfo{person}{Jingyu Liu}, \bibinfo{person}{Romain Sauvestre}, \bibinfo{person}{Tal Remez}, {et~al\mbox{.}}} \bibinfo{year}{2023}\natexlab{}.
\newblock \showarticletitle{Code llama: Open foundation models for code}.
\newblock \bibinfo{journal}{\emph{arXiv preprint arXiv:2308.12950}} (\bibinfo{year}{2023}).
\newblock


\bibitem[Team(2025)]%
        {qwq32b}
\bibfield{author}{\bibinfo{person}{Qwen Team}.} \bibinfo{year}{2025}\natexlab{}.
\newblock \bibinfo{title}{QwQ-32B: Embracing the Power of Reinforcement Learning}.
\newblock
\urldef\tempurl%
\url{https://qwenlm.github.io/blog/qwq-32b/}
\showURL{%
\tempurl}


\bibitem[Tong and Zhang(2024)]%
        {tong-zhang-2024-codejudge}
\bibfield{author}{\bibinfo{person}{Weixi Tong} {and} \bibinfo{person}{Tianyi Zhang}.} \bibinfo{year}{2024}\natexlab{}.
\newblock \showarticletitle{{C}ode{J}udge: Evaluating Code Generation with Large Language Models}. In \bibinfo{booktitle}{\emph{Proceedings of the 2024 Conference on Empirical Methods in Natural Language Processing}}, \bibfield{editor}{\bibinfo{person}{Yaser Al-Onaizan}, \bibinfo{person}{Mohit Bansal}, {and} \bibinfo{person}{Yun-Nung Chen}} (Eds.). \bibinfo{publisher}{Association for Computational Linguistics}, \bibinfo{address}{Miami, Florida, USA}, \bibinfo{pages}{20032--20051}.
\newblock
\href{https://doi.org/10.18653/v1/2024.emnlp-main.1118}{doi:\nolinkurl{10.18653/v1/2024.emnlp-main.1118}}


\bibitem[Wei et~al\mbox{.}(2025)]%
        {wei2025evaluating}
\bibfield{author}{\bibinfo{person}{Minnan Wei}, \bibinfo{person}{Ziming Li}, \bibinfo{person}{Xiang Chen}, \bibinfo{person}{Menglin Zheng}, \bibinfo{person}{Ziyan Qu}, \bibinfo{person}{Cheng Yu}, \bibinfo{person}{Siyu Chen}, {and} \bibinfo{person}{Xiaolin Ju}.} \bibinfo{year}{2025}\natexlab{}.
\newblock \showarticletitle{Evaluating and Improving Large Language Models for Competitive Program Generation}.
\newblock \bibinfo{journal}{\emph{arXiv preprint arXiv:2506.22954}} (\bibinfo{year}{2025}).
\newblock


\bibitem[Yang et~al\mbox{.}(2025)]%
        {qwen3}
\bibfield{author}{\bibinfo{person}{An Yang}, \bibinfo{person}{Anfeng Li}, \bibinfo{person}{Baosong Yang}, \bibinfo{person}{Beichen Zhang}, \bibinfo{person}{Binyuan Hui}, \bibinfo{person}{Bo Zheng}, \bibinfo{person}{Bowen Yu}, \bibinfo{person}{Chang Gao}, \bibinfo{person}{Chengen Huang}, \bibinfo{person}{Chenxu Lv}, \bibinfo{person}{Chujie Zheng}, \bibinfo{person}{Dayiheng Liu}, \bibinfo{person}{Fan Zhou}, \bibinfo{person}{Fei Huang}, \bibinfo{person}{Feng Hu}, \bibinfo{person}{Hao Ge}, \bibinfo{person}{Haoran Wei}, \bibinfo{person}{Huan Lin}, \bibinfo{person}{Jialong Tang}, \bibinfo{person}{Jian Yang}, \bibinfo{person}{Jianhong Tu}, \bibinfo{person}{Jianwei Zhang}, \bibinfo{person}{Jianxin Yang}, \bibinfo{person}{Jiaxi Yang}, \bibinfo{person}{Jing Zhou}, \bibinfo{person}{Jingren Zhou}, \bibinfo{person}{Junyang Lin}, \bibinfo{person}{Kai Dang}, \bibinfo{person}{Keqin Bao}, \bibinfo{person}{Kexin Yang}, \bibinfo{person}{Le Yu}, \bibinfo{person}{Lianghao Deng}, \bibinfo{person}{Mei Li}, \bibinfo{person}{Mingfeng Xue}, \bibinfo{person}{Mingze Li}, \bibinfo{person}{Pei Zhang}, \bibinfo{person}{Peng Wang}, \bibinfo{person}{Qin Zhu}, \bibinfo{person}{Rui Men}, \bibinfo{person}{Ruize Gao}, \bibinfo{person}{Shixuan Liu}, \bibinfo{person}{Shuang Luo}, \bibinfo{person}{Tianhao Li}, \bibinfo{person}{Tianyi Tang}, \bibinfo{person}{Wenbiao Yin}, \bibinfo{person}{Xingzhang Ren}, \bibinfo{person}{Xinyu Wang}, \bibinfo{person}{Xinyu Zhang}, \bibinfo{person}{Xuancheng Ren}, \bibinfo{person}{Yang Fan}, \bibinfo{person}{Yang Su}, \bibinfo{person}{Yichang Zhang}, \bibinfo{person}{Yinger Zhang}, \bibinfo{person}{Yu Wan}, \bibinfo{person}{Yuqiong Liu}, \bibinfo{person}{Zekun Wang}, \bibinfo{person}{Zeyu Cui}, \bibinfo{person}{Zhenru Zhang}, \bibinfo{person}{Zhipeng Zhou}, {and} \bibinfo{person}{Zihan Qiu}.} \bibinfo{year}{2025}\natexlab{}.
\newblock \showarticletitle{Qwen3 Technical Report}.
\newblock \bibinfo{journal}{\emph{arXiv preprint arXiv:2505.09388}} (\bibinfo{year}{2025}).
\newblock


\bibitem[Yu et~al\mbox{.}(2024)]%
        {yu2024codereval}
\bibfield{author}{\bibinfo{person}{Hao Yu}, \bibinfo{person}{Bo Shen}, \bibinfo{person}{Dezhi Ran}, \bibinfo{person}{Jiaxin Zhang}, \bibinfo{person}{Qi Zhang}, \bibinfo{person}{Yuchi Ma}, \bibinfo{person}{Guangtai Liang}, \bibinfo{person}{Ying Li}, \bibinfo{person}{Qianxiang Wang}, {and} \bibinfo{person}{Tao Xie}.} \bibinfo{year}{2024}\natexlab{}.
\newblock \showarticletitle{Codereval: A benchmark of pragmatic code generation with generative pre-trained models}. In \bibinfo{booktitle}{\emph{Proceedings of the 46th IEEE/ACM International Conference on Software Engineering}}. \bibinfo{pages}{1--12}.
\newblock


\bibitem[Zhang et~al\mbox{.}(2024)]%
        {zhang2024codeagentenhancingcodegeneration}
\bibfield{author}{\bibinfo{person}{Kechi Zhang}, \bibinfo{person}{Jia Li}, \bibinfo{person}{Ge Li}, \bibinfo{person}{Xianjie Shi}, {and} \bibinfo{person}{Zhi Jin}.} \bibinfo{year}{2024}\natexlab{}.
\newblock \bibinfo{title}{CodeAgent: Enhancing Code Generation with Tool-Integrated Agent Systems for Real-World Repo-level Coding Challenges}.
\newblock
\showeprint[arxiv]{2401.07339}~[cs.SE]
\urldef\tempurl%
\url{https://arxiv.org/abs/2401.07339}
\showURL{%
\tempurl}


\bibitem[Zhang et~al\mbox{.}(2025)]%
        {zhang2025llm}
\bibfield{author}{\bibinfo{person}{Ziyao Zhang}, \bibinfo{person}{Chong Wang}, \bibinfo{person}{Yanlin Wang}, \bibinfo{person}{Ensheng Shi}, \bibinfo{person}{Yuchi Ma}, \bibinfo{person}{Wanjun Zhong}, \bibinfo{person}{Jiachi Chen}, \bibinfo{person}{Mingzhi Mao}, {and} \bibinfo{person}{Zibin Zheng}.} \bibinfo{year}{2025}\natexlab{}.
\newblock \showarticletitle{Llm hallucinations in practical code generation: Phenomena, mechanism, and mitigation}.
\newblock \bibinfo{journal}{\emph{Proceedings of the ACM on Software Engineering}} \bibinfo{volume}{2}, \bibinfo{number}{ISSTA} (\bibinfo{year}{2025}), \bibinfo{pages}{481--503}.
\newblock


\bibitem[Zhao et~al\mbox{.}(2024)]%
        {zhao2024explainability}
\bibfield{author}{\bibinfo{person}{Haiyan Zhao}, \bibinfo{person}{Hanjie Chen}, \bibinfo{person}{Fan Yang}, \bibinfo{person}{Ninghao Liu}, \bibinfo{person}{Huiqi Deng}, \bibinfo{person}{Hengyi Cai}, \bibinfo{person}{Shuaiqiang Wang}, \bibinfo{person}{Dawei Yin}, {and} \bibinfo{person}{Mengnan Du}.} \bibinfo{year}{2024}\natexlab{}.
\newblock \showarticletitle{Explainability for large language models: A survey}.
\newblock \bibinfo{journal}{\emph{ACM Transactions on Intelligent Systems and Technology}} \bibinfo{volume}{15}, \bibinfo{number}{2} (\bibinfo{year}{2024}), \bibinfo{pages}{1--38}.
\newblock


\bibitem[Zheng et~al\mbox{.}(2024)]%
        {zheng2024opencodeinterpreter}
\bibfield{author}{\bibinfo{person}{Tianyu Zheng}, \bibinfo{person}{Ge Zhang}, \bibinfo{person}{Tianhao Shen}, \bibinfo{person}{Xueling Liu}, \bibinfo{person}{Bill~Yuchen Lin}, \bibinfo{person}{Jie Fu}, \bibinfo{person}{Wenhu Chen}, {and} \bibinfo{person}{Xiang Yue}.} \bibinfo{year}{2024}\natexlab{}.
\newblock \showarticletitle{Opencodeinterpreter: Integrating code generation with execution and refinement}.
\newblock \bibinfo{journal}{\emph{arXiv preprint arXiv:2402.14658}} (\bibinfo{year}{2024}).
\newblock


\end{thebibliography}

\end{document}